\documentclass[11pt]{article}
\usepackage{amsfonts}
\usepackage[margin=2cm]{geometry}
\usepackage{amsmath,amssymb,amsthm,array,booktabs,cite,dsfont,graphicx,mathtools,multirow,placeins,rotating,stmaryrd,tensor,verbatim}
\usepackage[bookmarksnumbered,linktocpage,pdfstartview=FitH]{hyperref}
\usepackage{hypcap}
\usepackage[margin=2cm]{geometry}
\usepackage{slashbox}
\usepackage{caption}
\usepackage{color}
\usepackage{setspace}
\usepackage{slashed}
\usepackage{changepage}

\newcolumntype{M}[1]{>{$}{#1}<{$}}

\newcommand{\sst}[1]{{\scriptscriptstyle #1}}

\newcommand{\rep}[1]{\ensuremath{\mathbf{#1}}}

\def\0{{\sst{(0)}}}
\def\1{{\sst{(1)}}}
\def\2{{\sst{(2)}}}
\def\3{{\sst{(3)}}}
\def\4{{\sst{(4)}}}
\def\5{{\sst{(5)}}}
\def\6{{\sst{(6)}}}
\def\7{{\sst{(7)}}}

\newcommand{\be}{\begin{equation}}
\newcommand{\ee}{\end{equation}}
\def\ba{\begin{array}}
\def\ea{\end{array}}

\newcommand{\bea}{\begin{eqnarray}}
\newcommand{\eea}{\end{eqnarray}}

\DeclareMathOperator{\tr}{tr}
\DeclareMathOperator{\End}{End}
\DeclareMathOperator{\SO}{SO}
\DeclareMathOperator{\Orth}{O}
\DeclareMathOperator{\USp}{USp}
\DeclareMathOperator{\SL}{SL}
\DeclareMathOperator{\SU}{SU}
\DeclareMathOperator{\Sp}{Sp}

\DeclareMathOperator{\Un}{U}

\DeclareMathOperator{\Cliff}{Cliff}
\DeclareMathOperator{\Pin}{Pin}
\DeclareMathOperator{\Spin}{Spin}

\newcommand{\J}{\mathfrak{J}}

\newcommand{\alg}{\mathds{A}}
\newcommand{\mf}{\mathfrak}

\newcommand{\R}{\mathds{R}}
\newcommand{\C}{\mathds{C}}
\newcommand{\Q}{\mathds{H}}
\newcommand{\Oct}{\mathds{O}}

\newcommand{\N}{\mathcal{N}}

\newcommand{\su}{\mathfrak{su}}

\newcommand{\usp}{\mathfrak{usp}}
\newcommand{\spa}{\mathfrak{sp}}

\begin{document}

\begin{titlepage}%1
\begin{center}
\hfill DIAS-STP-15-02\\
\vskip 2cm

{\huge \bf Global symmetries of Yang-Mills squared in various dimensions}

\vskip 1.5cm

{\bf A. Anastasiou${}^1$, L.~Borsten${}^{1,2}$, M.~J.~Hughes${}^1$, and
S.~Nagy${}^{1,3}$}

\vskip 20pt

{\it ${}^1$Theoretical Physics, Blackett Laboratory, Imperial College London,\\
 London SW7 2AZ, United Kingdom}\\\vskip 5pt
 {\it ${}^2$School of Theoretical Physics, Dublin Institute for Advanced Studies,\\
10 Burlington Road, Dublin 4, Ireland}\\\vskip 5pt
{\it ${}^3$Department of Mathematics, Instituto Superior T\'{e}cnico,\\
Av. Rovisco Pais, 1049-001 Lisbon, Portugal}\\\vskip 5pt
\texttt{alexandros.anastasiou07@imperial.ac.uk}\\
\texttt{leron@stp.dias.ie}\\
\texttt{mia.hughes07@imperial.ac.uk}\\
\texttt{snagy@math.tecnico.ulisboa.pt}

\end{center}

\vskip 2.2cm

\begin{center} {\bf ABSTRACT}\\[3ex]\end{center}

Tensoring two on-shell super Yang-Mills multiplets in dimensions $D\leq 10$ yields an on-shell  supergravity multiplet, possibly with additional matter multiplets. Associating a (direct sum of) division algebra(s) $\mathds{D}$ with each dimension $3\leq D\leq 10$ we obtain  a formula for the supergravity  U-duality  $G$ and its maximal compact subgroup $H$ in terms of the  internal global symmetry algebras  of each super Yang-Mills theory. We  extend our analysis to include supergravities coupled to an arbitrary number of matter multiplets by allowing for  non-supersymmetric multiplets in the tensor product.

%\end{center}

%\noindent

\vfill

\end{titlepage}

\newpage \setcounter{page}{1} \numberwithin{equation}{section} \tableofcontents 

\newpage

\section{Introduction}

The idea of understanding aspects of quantum gravity in terms of a  double-copy of gauge theories has a long history going back at least to the  Kawai-Lewellen-Tye relations of string theory \cite{Kawai:1985xq}. There has since been a wealth of  developments expanding on this concept, perhaps most notably, but certainly not exclusively, in the context of gravitational and gauge scattering amplitudes. See for example \cite{Siegel:1988qu, Antoniadis:1992sa,Siegel:1995px,Sen:1995ff, Bianchi:2008pu, Bern:2008qj,Bern:2010ue,Bern:2010yg,Bern:2009kd, Katsaroumpas:2009iy, BjerrumBohr:2010rt, Hodges:2011wm, Damgaard:2012fb, Carrasco:2012ca, Huang:2012wr,  Boels:2013bi, Carrasco:2013ypa, Elvang:2013cua, Bern:2013uka, Cachazo:2013iea,Mason:2013sva, Dolan:2013isa, Chiodaroli:2014xia, Geyer:2014fka, Anastasiou:2014qba,Monteiro:2014cda}. Indeed, invoking the Bern-Carrasco-Johansson colour-kinematic duality it has been  conjectured  \cite{Bern:2010ue} that the on-mass-shell momentum-space scattering amplitudes for gravity are the  ``double-copy''  of gluon scattering amplitudes in Yang-Mills theory to all orders in perturbation theory.

This remarkable and somewhat surprising proposal  motivates the question: to what extent can one regard quantum gravity  as the double copy of Yang-Mills theory? In this context it is natural to ask how the symmetries of each theory are related. In recent work \cite{Anastasiou:2014qba} it was shown that the  off-shell \emph{local} transformation rules of  (super)gravity (namely general covariance, local Lorentz invariance, $p$-form gauge invariance and local supersymmetry) may be derived from those of flat space Yang-Mills (namely local gauge invariance and global super-Poincare) at the linearised level.

 Equally  important in the context of M-theory are the non-compact \emph{global} symmetries of supergravity \cite{Cremmer:1979up}, which are intimately related to the concept of U-duality \cite{Duff:1990hn, Hull:1994ys}.   For previous work on global symmetries in $D=4$ spacetime dimensions via squaring see \cite{Bianchi:2008pu,Damgaard:2012fb, Carrasco:2012ca,Chiodaroli:2014xia}. It was  shown in  \cite{Borsten:2013bp} that tensoring two   $D=3, \,\mathcal{N}=1,2,4,8$ super Yang-Mills  mulitplets results in a ``Freudenthal magic square  of supergravity  theories'', as summarised in \autoref{tab:3Dsugra}. 
The corresponding  Lie algebras  of \autoref{tab:3Dsugra}   are concisely summarised  by the  magic square formula \cite{Barton:2003, Borsten:2013bp}, 
\be\label{eq:tri}
\mf{L}_3(\alg_{ \mathcal{N}_L}, \alg_{ \mathcal{N}_R}):=\mf{tri}(\alg_{ \mathcal{N}_L})\oplus \mf{tri}(\alg_{ \mathcal{N}_R})+3(\alg_{ \mathcal{N}_L}\otimes\alg_{ \mathcal{N}_R}),
\ee
which takes as its argument a pair of division algebras $\alg_{\mathcal{N}_L}, \alg_{ \mathcal{N}_R}=\R, \C, \Q, \Oct$, where we have adopted the convention that $\dim \alg_{\mathcal{N}}=\mathcal{N}$.  The triality  algebra of $\alg$, denoted $\mf{tri}(\alg)$, is related to the total  on-shell global symmetries of the associated super Yang-Mills theory \cite{Anastasiou:2013cya}. This rather surprising connection, relating the magic square of Lie algebras to the square of super Yang-Mills, can be attributed to  the existence of a unified $\alg_\mathcal{N}=\R, \C, \Q, \Oct$ description of $D=3,\, \mathcal{N}=1,2,4,8$ super Yang-Mills theories.

\begin{table}[h]
 \begin{center}
\scriptsize
 \begin{tabular}{c|llllllllllll}
 \hline
 \hline
 \\
$\alg_{\N_L}\backslash\alg_{\N_R}$&&$\R$&$\C$&$\Q$&$\Oct$\\
\\
\hline
\\
&&$\mathcal{N}=2, f=4$&$\mathcal{N}=3, f=8$&$\mathcal{N}=5, f=16$&$\mathcal{N}=9, f=32$\\
$\R$&&$G=\SL(2,\R)$&$G=\SU(2,1)$&$G=\USp(4,2)$&$G=F_{4(-20)}$\\
&&$H=\SO(2)$&$H=\SO(3) \times \SO(2)$&$H=\SO(5)\times \SO(3)$&$H=\SO(9)$\\
&&&&\\
&&$\mathcal{N}=3, f=8$&$\mathcal{N}=4, f=16$&$\mathcal{N}=6, f=32$&$\mathcal{N}=10, f=64$\\
$\C$&&$G=\SU(2,1)$&$G=\SU(2,1)^2$&$G=\SU(4,2)$&$G=E_{6(-14)}$\\
&&$H=\SO(3) \times \SO(2)$&$H=\SO(3)^2 \times \SO(2)^2$&$H=\SO(6) \times \SO(3)\times \SO(2)$&$H=\SO(10) \times \SO(2)$\\
&&&\\
&&$\mathcal{N}=5, f=16$&$\mathcal{N}=6, f=32$&$\mathcal{N}=8, f=64$&$\mathcal{N}=12, f=128$\\
$\Q$&&$G=\USp(4,2)$&$G=\SU(4,2)$&$G=\SO(8,4)$&$G=E_{7(-5)}$\\
&&$H=\SO(5)\times \SO(3)$&$H=\SO(6)\times \SO(3)\times \SO(2)$&$H=\SO(8)\times \SO(3)\times \SO(3)$&$H=\SO(12)\times \SO(3)$\\
&&&&\\
&&$\mathcal{N}=9, f=32$&$\mathcal{N}=10, f=64$&$\mathcal{N}=12, f=128$&$\mathcal{N}=16, f=256$\\
$\Oct$&&$G=F_{4(-20)}$&$G=E_{6(-14)}$&$G=E_{7(-5)}$&$G=E_{8(8)}$\\
&&$H=\SO(9)$&$H=\SO(10) \times \SO(2)$&$H=\SO(12)\times \SO(3)$&$H=\SO(16)$\\
\\
\hline
\hline
\end{tabular}
\caption{\footnotesize{$(\N=\N_L+\N_R)$-extended $D=3$ supergravities obtained by left/right super Yang-Mills multiplets with $\mathcal{N}_L, \mathcal{N}_R=1,2,4,8$.  The 
algebras of  the corresponding U-duality  groups $G$ and their maximal compact subgroups $H$ are given by the  magic square  of Freudenthal-Rosenfeld-Tits \cite{Freudenthal:1954,Tits:1955, Freudenthal:1959,Rosenfeld:1956, Tits:1966}.  $f$ denotes the total number of degrees of freedom  in the resulting supergravity and matter  multiplets.}}
\label{tab:3Dsugra}
  \end{center}
\end{table}

 This observation was subsequently  generalised  to  $D=3, 4, 6$ and $10$ dimensions \cite{Anastasiou:2013cya, Anastasiou:2013hba} by incorporating  the well-known relationship  between the existence of minimal super Yang-Mills theories in  $D=3,4,6,10$ and the existence of  the four division algebras $\R, \C, \Q, \Oct$ \cite{Hurwitz:1898, Kugo:1982bn, Sudbery:1984, Baez:2001dm, Baez:2009xt}. From this perspective the $D=3$ magic square forms the base of  a ``magic pyramid" of supergravities. 
 
 These constructions build on a long line of work relating  division algebras and magic squares to spacetime and supersymmetry. See \cite{Gunaydin:1975mp, Gunaydin:1976vq,Gursey:1978et, Gunaydin:1979df, Julia:1980gr, Kugo:1982bn, Gunaydin:1983rk, Gunaydin:1983bi, Gunaydin:1984ak, Sudbery:1984, Sierra:1986dx, Gursey:1987mv,  Green:1987sp, Evans:1987tm, Duff:1987qa,Blencowe:1988sk, Gunaydin:1992zh, Berkovits:1993hx,  Manogue:1993ja, Evans:1994cn, Schray:1994ur, gursey1996role, Manogue:1998rv, Cremmer:1999du, Gunaydin:2000xr, Baez:2001dm,Toppan:2003yx,  Gunaydin:2005zz, Kuznetsova:2006ws, Bellucci:2006xz, Borsten:2008wd, Borsten:2009zy, Baez:2009xt, Baez:2010ye,Borsten:2010aa, Gunaydin:2010fi, Rios:2011fa, Huerta:2011ic, Huerta:2011aa, Ferrara:2011xb, Cacciatori:2012sf, Cacciatori:2012cb, Anastasiou:2014zfa, Huerta:2014loa, Marrani:2014qia} for a  glimpse  of the relevant  literature.  An early example\footnote{As far as we are aware the first instance in this context.}, closely related to the present contribution, appears in work the Julia \cite{Julia:1980gr} on  group disintergrations in supergravity. The oxidation of $\mathcal{N}$-extended  $D=3$ dimensional supergravity theories yields a partially symmetric ``trapezoid'' of non-compact global symmetries for $D=3,4,\ldots 11$ and $0, 2^0, 2^1,\ldots 2^7$ supercharges\footnote{It also includes the affine Kac-Moody algebras, $\mathfrak{e}_{9}=\mathfrak{e}_{8}^{+}, \mathfrak{e}_{7}^{+}, \mathfrak{e}_{6}^{+}, \mathfrak{so}_{10}^{+}$ in $D=2$ as made more precise in \cite{Julia:1981wc}.}. The subset of algebras in the trapezoid given by $D=3,4,5$ and $2^5, 2^6, 2^7$ supercharges fits into the $3\times 3$ inner $\C, \Q, \Oct$ part of the magic  square, excluding the $(\C, \C)$ entry.  Note, the exact symmetry of this subsquare is broken by the precise set of real forms obtained, which 
 are not given by any magic square formula in the conventional sense\footnote{We thank Benard Julia for bringing this observation, emphasised in \cite{Julia:1982gx}, to our attention.}. However, this set of theories also matches  the $D=3,4,5$ exterior wall of the pyramid in \autoref{fig:Galldims}  obtained  by squaring Yang-Mills and the corresponding algebras are indeed given by the pyramid formula \eqref{eq:Galg} described in this work.  Note, dispensing with the requirement of supersymmetry the same $3\times 3$ square but  with maximally non-compact real forms was derived as a corner of a ``magic triangle'' of theories in $3\leq D\leq 11$ spacetime dimensions \cite{Cremmer:1999du}. The entries of triangle are parametrised by the dimension $D$  of the theory and the rank $0\leq n\leq 8$ of its symmetry algebra. The complete magic triangle  displays  a remarkable symmetry under   $D \rightarrow 11-n, n\rightarrow 11-D$. It should be noted that these are not the only  magic triangles of Lie algebras, a particularly elegant and intriguing example being that of Cvitanovi\'c \cite{cvitanovic1977classical, cvitanovic2008group}.

 Returning to the theme of gravity as the square of Yang-Mills, the magic pyramid of \cite{Anastasiou:2013hba} corresponds to  a rather special subset of supergravity theories: those  given by tensoring the  $D=3,4,6,10$ division algebraic super Yang-Mills theories   constructed in \cite{Anastasiou:2013cya}.  In the present work we address the natural question of  generalisation beyond this select subclass of theories and give a new pyramid formula which makes the double-copy structure manifest for all possible products of supersymmetric Yang-Mills theories.
 
 In \autoref{sec:formulae} we consider all tensor products of left $\N_L$-extended and right $\N_R$-extended super Yang-Mills multiplets in $D=3,\ldots, 10$ dimensions and introduce three  formulae describing the global symmetries of the resulting $(\mathcal{N}_L+\N_R)$-extended supergravity  multiplets:
 \begin{enumerate}
 \item The algebra $\mathfrak{ra}(\mathcal{N}_L+\N_R, {D})$ of $(\mathcal{N}_L+\N_R)$-extended R-symmetry in $D$ dimensions,
 \be\label{eq:Rsym}
\mathfrak{ra}(\mathcal{N}_L+\N_R, {D})=\mathfrak{a}(\mathcal{N}_L, \mathds{D})\oplus\mathfrak{a}(\mathcal{N}_R,\mathds{D})+\mathds{D}[\mathcal{N}_L, \mathcal{N}_R];
\ee 
 \item The algebra $\mathfrak{h}(\mathcal{N}_L+\mathcal{N}_R, D)$ of $H$, the maximal compact subgroup of the U-duality group $G$,   
    \be\label{eq:Halg}
\mathfrak{h}(\mathcal{N}_L+\mathcal{N}_R, D)=\mathfrak{int}(\mathcal{N}_L,D)\oplus\mathfrak{int}(\mathcal{N}_R,D)\oplus\delta_{D, 4}\mathfrak{u}(1)+\mathds{D}[\mathcal{N}_L, \mathcal{N}_R];
\ee 
 \item The algebra $\mathfrak{g}(\mathcal{N}_L+\mathcal{N}_R, D)$ of the U-duality group $G$ itself,
 \be\label{eq:Galg}
\mathfrak{g}(\mathcal{N}_L+\mathcal{N}_R,D)
=\mathfrak{h}(\mathcal{N}_L+\mathcal{N}_R,D)+\mathds{D}_{*}[\mathcal{N}_L]\otimes\mathds{D}_{*}[\mathcal{N}_R]+\mathds{D}[\mathcal{N}_L, \mathcal{N}_R]+\R_L\otimes\R_R+i\delta_{D,4}\R_L\otimes\R_R.
\ee 
 \end{enumerate}
Here we have used $\oplus$ and $+$ to distinguish the direct sum between Lie algebras and vector spaces; only if $[\mf{m}, \mf{n}]=0$ do we use $\mf{m}\oplus\mf{n}$. The  meaning of these formulae and, in particular, their relation to the symmetries of the left and right super Yang-Mills factors, will be described in \autoref{sec:formulae}. For the moment we simply note  that they make the  left/right  structure manifest and uniform for all $\mathcal{N}_L, \mathcal{N}_R$ and $D$ and, as we shall see, each  summand appearing in the three formulae has a natural $left \otimes right$ origin. The groups $H$ and $G$ corresponding to  \eqref{eq:Halg} and \eqref{eq:Galg} are given in the generalised pyramids of \autoref{fig:Halldims} and \autoref{fig:Galldims}, respectively.  For these groups, the formulae presented above can be regarded as generalised  ``matrix models'', in the sense of \cite{Barton:2003} (not to be confused with  (M)atrix models), for classical and  exceptional  Lie algebras. As a matrix model, it is perhaps not as elegant as those presented in \cite{Barton:2003}. For one, we make \emph{no} use of the octonions. However, it has the advantage, from our perspective,  that it describes systematically all  groups obtained by squaring super Yang-Mills and, moreover, makes the left and right factors manifest.
  
For $\N_L+\N_R$ half-maximal or less the super Yang-Mills tensor products yield  supergravity multiplets together with additional matter multiplets, as described in \autoref{tab:tensors1}. They may always be obtained as consistent truncations or, in many cases, factorised orbifold truncations of the maximally supersymmetric cases, as in \cite{Chiodaroli:2014xia}. The type, number and coupling of these multiplets is fixed with respect to   \eqref{eq:Halg} and \eqref{eq:Galg}. However, as we shall describe in \autoref{sec:GeneralisedTensorProducts}, by including a non-supersymmetric factor in the tensor product these matter couplings may be generalised to include an arbitrary number of vector multiplets (thus  clearly not truncations). This procedure  naturally yields analogous formulae for $\mathfrak{h}$ and $\mathfrak{g}$, corresponding to specific couplings. The nature of these couplings is in a certain sense as simple as possible. This follows from the  symmetries, which may be regarded as a consequence of simple interactions, assumed to be present in the non-supersymmetric factor of the tensor product.

\section{Global symmetries of super Yang-Mills squared}\label{sec:formulae}

\subsection{Tensoring super Yang-Mills theories in $D\geq 3$}

Tensoring   $\N_L$-extended and  $\N_R$-extended super Yang-Mills multiplets, $[\mathcal{N}_{L}]_{V}$ and $[\mathcal{N}_{R}]_{V}$, yields an $(\N_L+\N_R)$-extended supergravity multiplet, $[\mathcal{N}_L+\mathcal{N}_R]_{grav}$,
 \be
[\mathcal{N}_L]_{V}\otimes[\mathcal{N}_R]_{V}\to[\mathcal{N}_L+\mathcal{N}_R]_{grav}+[\mathcal{N}_L+\mathcal{N}_R]_{matter},
\ee
with additional matter multiplets, $[\mathcal{N}_L+\mathcal{N}_R]_{matter}$, for $[\N_L+\N_R]_{grav}$ half-maximal or less. See  \autoref{tab:tensors1} and \autoref{tab:tensors2}.

We consider on-shell space-time little group super Yang-Mills multiplets with global symmetry algebra 
\be
\mathfrak{so}(D-2)_{ST}\oplus\mathfrak{int}(\N, D),
\ee where $\mathfrak{int}(\N, D)$ denotes the global internal symmetry algebra of the Lagrangian. 
For $\mathfrak{so}(D-2)_{ST}$ the tensor products are $\mathfrak{so}(D-2)_{ST}$-modules, while for $\mathfrak{int}(\N_L, D)$ and $\mathfrak{int}(\N_R, D)$ they are  $\mathfrak{int}(\N_L, D)\oplus\mathfrak{int}(\N_R, D)$-modules. Very schematically, the general tensor product is given by,
\be\label{eq:roughsquare}
\begin{array}{c|ccccc}
\otimes&\tilde{A}_{\nu}&\tilde{\lambda}^{a'} & \tilde{\phi}^{i'}\\
\hline
A_{\mu}&g_{\mu\nu}+B_{\mu\nu}+\phi&\psi_{\mu}^{a'}+\lambda^{a'}& A_{\mu}^{i'}\\
\lambda^{a}&\psi_{\nu}^{a}+\lambda^{a}&\phi_{RR}^{aa'}+\cdots&\lambda^{ai'}\\
 \phi^{i}&A_{\nu}^{i}&\lambda^{ia'}&\phi^{ii'}\\
\end{array}
\ee 
where $a, i$ and $a', i'$ are indices of the appropriate $\mathfrak{int}(\N_{L}, D)$ and $\mathfrak{int}(\N_{R}, D)$ representations, respectively. Note, we will always dualise $p$-forms to their lowest possible rank consistent with their little group representations, for example, $B_{\mu\nu}\rightarrow \phi, A_\mu$ in $D=4,5$, respectively. This ensures  U-duality  is manifest.  The particular set of Ramond-Ramond $p$-forms $\phi_{RR}^{aa'}+\cdots$ one obtains is dimension dependent.

The detailed form of these tensor products for $D>3$ are summarised in \autoref{tab:tensors1} and \autoref{tab:tensors2}, where for a given little group representation we have collected the  $\mathfrak{int}(\N_L, D)\oplus\mathfrak{int}(\N_R, D)$ representations into the appropriate representations of $\mathfrak{h}(\N_L+\N_R, D)$. For example, consider the square of the $D=5, \N=2$ super Yang-Mills multiplet, which has global symmetry algebra $\mathfrak{so}(3)_{ST}\oplus\mathfrak{sp}(2)$,
\be\label{eq:D5ex}
\begin{array}{c|ccccccc}
\otimes
&
 \begin{array}{c} \tilde{A}_\mu\\(\rep{3}; \rep{1})\end{array}& \begin{array}{c} \tilde{\lambda}\\(\rep{2}; \rep{4})\end{array}& \begin{array}{c} \tilde{\phi} \\(\rep{1}; \rep{5})\end{array}\\
\hline
A_\mu\;\;(\rep{3}; \rep{1}) &(\rep{5}; \rep{1,1})+ (\rep{3}; \rep{1,1})+(\rep{1}; \rep{1,1})&(\rep{4}; \rep{1,4})+(\rep{2}; \rep{1,4})&(\rep{3}; \rep{1,5})\\
\lambda\;\;(\rep{2}; \rep{4})&(\rep{4}; \rep{4,1})+(\rep{2}; \rep{4,1})& (\rep{3}; \rep{4,4})+(\rep{1}; \rep{4,4})&(\rep{2}; \rep{4,5})\\
 \phi \;\; (\rep{1}; \rep{5})&(\rep{3}; \rep{5,1})&(\rep{2}; \rep{5,4})&(\rep{1}; \rep{5,5})
 \end{array}
\ee

\newcolumntype{M}{>{$}c<{$}}
\begin{table}[h]
\tiny

\hspace{-0.5in}\begin{tabular}{M||M|M|M }
 \hline
 \hline
 &&\\
D=4, \mathfrak{u}(1)_{ST} &  
\begin{array}{ccccccc} \multicolumn{3}{c}{\N=4\;\; \mathfrak{su}(4)} \\ [3pt] \begin{array}{c}A_\mu\\(1; \rep{1})\\+c.c.\end{array}&\begin{array}{c}\lambda\\(\frac{1}{2}; \rep{4})\\+c.c.\end{array}& \begin{array}{c} \phi\\(0; \rep{6})\\\end{array}\end{array}&
\begin{array}{ccccccc} \multicolumn{3}{c}{\N=2 \;\;  \mathfrak{u}(2)} \\ [3pt] \begin{array}{c}A_\mu\\(1; \rep{1}(0))\\+c.c.\end{array}&  \begin{array}{c}\lambda\\(\frac{1}{2}; \rep{2}(1))\\+c.c.\end{array}&  \begin{array}{c}\phi\\(0; \rep{1}(2))\\+c.c.\end{array}\end{array}&
\begin{array}{cc} \multicolumn{2}{c}{\N=1\;\;  \mathfrak{u}(1)} \\[3pt]  \begin{array}{c}A_\mu\\(1;0)\\+c.c.\end{array}& \begin{array}{c}\lambda\\(\frac{1}{2}; 1)\\+c.c.\end{array}\end{array} \\
&&\\
\hline
\hline
&&\\
\begin{array}{cc}\N=4& \mathfrak{su}(4)  \\[3pt] A_\mu&(1; \rep{1})+c.c.\\\lambda&(\frac{1}{2}; \rep{4})+c.c.\\ [3pt]  \phi &(0; \rep{6})\end{array}&
\begin{array}{cc}\N=8& \mathfrak{su}(8) \\ [3pt]g_{\mu\nu}&(2; \rep{1})+c.c. \\[3pt]\psi_{\mu}&(\frac{3}{2}; \rep{8})+c.c.\\ [3pt]A_\mu&(1; \rep{28})+c.c. \\[3pt] \lambda&(\frac{1}{2}; \rep{56})+c.c. \\[3pt]  \phi &(0; \rep{70})\end{array}&
\begin{array}{cc}\N=6& \mathfrak{u}(6) \\[3pt]  g_{\mu\nu}&(2; \rep{1}(0))+c.c. \\[3pt] \psi_{\mu}&(\frac{3}{2}; \rep{{6}}(1))+c.c. \\[3pt]  A_\mu&(1; \rep{1}(-6)+\rep{{15}}(2))+c.c. \\[3pt] \lambda&(\frac{1}{2}; \rep{{6}}(-5)+\rep{20}(3))+c.c. \\[3pt]  \phi &(0; \rep{15}(-4))+c.c.\end{array} &
\begin{array}{cc}\N=5& \mathfrak{u}(5) \\ [3pt]g_{\mu\nu}&(2; \rep{1}(0))+c.c. \\[3pt]\psi_{\mu}&(\frac{3}{2}; \rep{{5}}(1))+c.c.\\ [3pt]A_\mu&(1; \rep{{10}}(2))+c.c. \\[3pt] \lambda&(\frac{1}{2}; \rep{1}(-5)+\rep{\overline{10}}(3))+c.c.\\  [3pt]\phi &(0; \rep{{5}}(-4))+c.c.\end{array} \\
&&\\
\hline
&&\\
\begin{array}{cc}\N=2& \mathfrak{u}(2) \\ [3pt] A_\mu&(1; \rep{1}(0))+c.c. \\[3pt] \lambda&(\frac{1}{2}; \rep{2}(1))+c.c. \\[3pt]  \phi &(0; \rep{1}(2))+c.c.\end{array}
&
\begin{array}{cc}\N=6& \mathfrak{u}(6) \\[3pt]  g_{\mu\nu}&(2; \rep{1}(0))+c.c. \\[3pt] \psi_{\mu}&(\frac{3}{2}; \rep{{6}}(1))+c.c. \\[3pt]  A_\mu&(1; \rep{1}(-6)+\rep{{15}}(2))+c.c. \\[3pt] \lambda&(\frac{1}{2}; \rep{{6}}(-5)+\rep{20}(3))+c.c. \\[3pt]  \phi &(0; \rep{15}(-4))+c.c.\end{array} &

 \begin{array}{cc}\N=4& \mathfrak{u}(4)\oplus\mathfrak{u}(1) \\ [3pt]g_{\mu\nu}&(2; \rep{1}(0)(0))+c.c. \\[3pt]\psi_{\mu}&(\frac{3}{2}; \rep{{4}}(1)(0))+c.c.\\ [3pt]A_\mu&(1; \rep{6}(2)(0))+c.c. \\[3pt] \lambda&(\frac{1}{2}; \rep{\overline{4}}(3)(0))+c.c. \\[3pt]  \phi &(0; \rep{1}(4)(0))+c.c.\\ [3pt] 2 [\N=4]_{V} &\\ [3pt] 2 \times A_\mu& \begin{array}{l}(1; \rep{1}(-2)(2))+c.c.\\(1; \rep{1}(-2)(-2))+c.c.\end{array} \\[3pt] 2 \times \lambda& \begin{array}{l} (\frac{1}{2}; \rep{4}(-1)(2))+c.c. \\(\frac{1}{2}; \rep{4}(-1)(-2))+c.c.\end{array} \\[3pt] 2 \times \phi & \begin{array}{l} (0; \rep{6}(0)(2))\\ (0; \rep{6}(0)(-2))\end{array}\\
 \end{array} &
 \begin{array}{cc}\N=3& \mathfrak{u}(3)\oplus \mathfrak{u}(1)\\ [3pt]g_{\mu\nu}&(2; \rep{1}(0)(0))+c.c. \\[3pt]\psi_{\mu}&(\frac{3}{2}; \rep{3}(1)(0))+c.c.\\ [3pt]A_\mu&(1; \rep{\overline{3}}(2)(0))+c.c. \\[3pt] \lambda&(\frac{1}{2}; \rep{1}(3)(0))+c.c.  \\ [3pt]  [\N=3]_{V} &\\ [3pt]  A_\mu& (1; \rep{1}(-2)(2))+c.c. \\[3pt] \lambda&  (\frac{1}{2}; \rep{3}(-1)(2)+\rep{1}(-1)(-2))+c.c. \\[3pt]   \phi &(0; \rep{\overline{3}}(0)(2))+c.c.\\
 \end{array}   \\
&&\\
\hline
&&\\
\begin{array}{cc}\N=1& \mathfrak{u}(1)  \\[3pt] A_\mu& (1;0)+c.c. \\[3pt] \lambda&(\frac{1}{2}; 1)+c.c.\\&\end{array} &
\begin{array}{cc}\N=5& \mathfrak{u}(5) \\ [3pt]g_{\mu\nu}&(2; \rep{1}(0))+c.c. \\[3pt]\psi_{\mu}&(\frac{3}{2}; \rep{{5}}(1))+c.c.\\ [3pt]A_\mu&(1; \rep{{10}}(2))+c.c. \\[3pt] \lambda&(\frac{1}{2}; \rep{1}(-5)+\rep{\overline{10}}(3))+c.c.\\  [3pt]\phi &(0; \rep{{5}}(-4))+c.c.\end{array} 
&
 \begin{array}{cc}\N=3& \mathfrak{u}(3)\oplus \mathfrak{u}(1)\\ [3pt]g_{\mu\nu}&(2; \rep{1}(0)(0))+c.c. \\[3pt]\psi_{\mu}&(\frac{3}{2}; \rep{3}(1)(0))+c.c.\\ [3pt]A_\mu&(1; \rep{\overline{3}}(2)(0))+c.c. \\[3pt] \lambda&(\frac{1}{2}; \rep{1}(3)(0))+c.c.  \\ [3pt]  [\N=3]_{V} &\\ [3pt]  A_\mu& (1; \rep{1}(-2)(2))+c.c. \\[3pt] \lambda&  (\frac{1}{2}; \rep{3}(-1)(2)+\rep{1}(-1)(-2))+c.c. \\[3pt]   \phi &(0; \rep{\overline{3}}(0)(2))+c.c.\\
\end{array}
 &
   \begin{array}{cc}\N=2& \mathfrak{u}(2)\oplus \mathfrak{u}(1)\\ [3pt]g_{\mu\nu}&(2; \rep{1}(0)(0))+c.c. \\[3pt]\psi_{\mu}&(\frac{3}{2}; \rep{2}(1)(0))+c.c.\\ [3pt]A_\mu&(1; \rep{1}(2)(0))+c.c. \\[3pt]    [\N=2]_{h} &\\ [3pt]  \lambda&  (\frac{1}{2}; \rep{1}(-3)(-1))+\rep{1}(1)(1))+c.c. \\[3pt]   \phi &(0; \rep{2}(-2)(-1))+c.c.\\\end{array}  \\
&&\\
 \hline
 \hline
 &&\\
D=5, \mathfrak{sp}(1)_{ST}
&
\begin{array}{ccc}\multicolumn{3}{c}{\N=2\;\;  \mathfrak{sp}(2)} \\ [3pt] \begin{array}{c} A_\mu\\(\rep{3}; \rep{1})\end{array}& \begin{array}{c} \lambda\\(\rep{2}; \rep{4})\end{array}& \begin{array}{c} \phi \\(\rep{1}; \rep{5})\end{array}\end{array}
&
\begin{array}{ccc}\multicolumn{3}{c}{\N=1\;\;  \mathfrak{sp}(1)} \\ [3pt] \begin{array}{c} A_\mu\\(\rep{3}; \rep{1})\end{array}& \begin{array}{c} \lambda\\(\rep{2}; \rep{2})\end{array}&  \begin{array}{c} \phi \\ (\rep{1}; \rep{1})\end{array}\end{array}\\
&&\\
\hline
\hline
&&\\
\begin{array}{cc}\N=2& \mathfrak{sp}(2) \\ [3pt] A_\mu&(\rep{3}; \rep{1})\\[3pt] \lambda&(\rep{2}; \rep{4})\\[3pt]  \phi &(\rep{1}; \rep{5})\end{array}
&
\begin{array}{cc}\N=4& \mathfrak{sp}(4) \\ [3pt]g_{\mu\nu}&(\rep{5}; \rep{1})\\[3pt]\psi_{\mu}&(\rep{4}; \rep{8})\\ [3pt]A_\mu&(\rep{3}; \rep{27})\\[3pt] \lambda&(\rep{2}; \rep{48})\\  [3pt]\phi &(\rep{1}; \rep{42})\end{array}  
&
\begin{array}{cc}\N=3& \mathfrak{sp}(3) \\ [3pt]g_{\mu\nu}&(\rep{5}; \rep{1})\\[3pt]\psi_{\mu}&(\rep{4}; \rep{6})\\ [3pt]A_\mu&(\rep{3}; \rep{1}+ \rep{14})\\[3pt] \lambda&(\rep{2}; \rep{6}+\rep{14'})\\  [3pt]\phi &(\rep{1}; \rep{14})\end{array} \\
&&\\
\hline
&&\\
\begin{array}{cc}\N=1& \mathfrak{sp}(1) \\ [3pt] A_\mu&(\rep{3}; \rep{1})\\[3pt] \lambda&(\rep{2}; \rep{2})\\[3pt]  \phi &(\rep{1}; \rep{1})\end{array}
&
\begin{array}{cc}\N=3& \mathfrak{sp}(3) \\ [3pt]g_{\mu\nu}&(\rep{5}; \rep{1})\\[3pt]\psi_{\mu}&(\rep{4}; \rep{6})\\ [3pt]A_\mu&(\rep{3}; \rep{1}+ \rep{14})\\[3pt] \lambda&(\rep{2}; \rep{6}+\rep{14'})\\  [3pt]\phi &(\rep{1}; \rep{14})\end{array} 
& 
\begin{array}{cc}\N=2& \mathfrak{sp}(2) \\ [3pt]g_{\mu\nu}&(\rep{5}; \rep{1})\\[3pt]\psi_{\mu}&(\rep{4}; \rep{4})\\ [3pt]A_\mu&(\rep{3}; \rep{1}+ \rep{5})\\[3pt] \lambda&(\rep{2}; \rep{4})\\  [3pt]\phi &(\rep{1}; \rep{1})\\[3pt] [\N=2]_V&  \\ [3pt] A_\mu&(\rep{3}; \rep{1})\\[3pt] \lambda&(\rep{2}; \rep{4})\\[3pt]  \phi &(\rep{1}; \rep{5})\end{array}  \\
&&\\
\hline
\hline
\end{tabular}
\caption{\footnotesize Tensor products of left and right super Yang-Mills multiplets in $D=4,5$. Dimensions $D=6,7,8,9,10$ are given in \autoref{tab:tensors2}.  In $(\mathbf{m; n})$ $\mathbf{m}$ denotes the spacetime little group representation and $\mathbf{n}$ the representation of the internal global symmetry displayed, $\mathfrak{int}$ for the super Yang-Mills multiplets and $\mathfrak{h}$ for the resulting supergravity $+$ matter multiplets. Here $V$ and $h$ denote vector and hyper multiplets, respectively.} \label{tab:tensors1}
\end{table}
 \newcolumntype{M}{>{$}c<{$}}
\begin{table}[h]
\tiny
\centering
\begin{tabular}{M||M|M }
 \hline
 \hline
 &&\\
 D=6, \mathfrak{sp}(1)\oplus\mathfrak{sp}(1)_{ST}
 &
 \begin{array}{cccccc}\multicolumn{3}{c}{\N=(1,1)\;\; \mathfrak{sp}(1)\oplus\mathfrak{sp}(1)} \\ [3pt] A_\mu\;(\rep{2},\rep{2}; \rep{1,1})& \lambda\;(\rep{2, 1; \rep{1,2})+(\rep{1,2}; \rep{2,1})}& \phi \;(\rep{1, 1}; \rep{2,2})\\
\end{array}
 &
 \begin{array}{cc}\multicolumn{2}{c}{\N=(1,0)\;\; \mathfrak{sp}(1)\oplus\varnothing} \\ [3pt] A_\mu\;(\rep{2},\rep{2}; \rep{1})& \lambda\;(\rep{1,2}; \rep{2})\\
\end{array}
\\
&&\\
\hline
\hline
&&\\
\begin{array}{cc}\N=(1,1)& \mathfrak{sp}(1)\oplus\mathfrak{sp}(1) \\ [3pt] A_\mu&(\rep{2},\rep{2}; \rep{1})\\[3pt] \lambda&(\rep{2, 1; \rep{1,2})+(\rep{1,2}; \rep{2,1})}\\[3pt] \phi &(\rep{1, 1}; \rep{2,2})\\
\end{array}
&
\begin{array}{cc}\N=(2,2)&\mathfrak{sp}(2)\oplus\mathfrak{sp}(2) \\ [3pt] g_{\mu\nu}&(\rep{3},\rep{3}; \rep{1,1})\\[3pt] \psi_\mu&(\rep{2},\rep{3}; \rep{4,1})+(\rep{3},\rep{2}; \rep{1,4})\\[3pt] A_\mu&(\rep{2},\rep{2}; \rep{4,4})\\[3pt] A_{\mu\nu}&(\rep{3},\rep{1}; \rep{1, 5})+(\rep{1},\rep{3}; \rep{5, 1})\\[3pt]  \lambda&(\rep{2,1}; \rep{4,5})+(\rep{1,2}; \rep{5,4})\\[3pt]  \phi &(\rep{1,1}; \rep{5,5})\\
\end{array}
&
\begin{array}{cc}\N=(2,1)&\mathfrak{sp}(2)\oplus\mathfrak{sp}(1) \\ [3pt] g_{\mu\nu}&(\rep{3},\rep{3}; \rep{1,1})\\[3pt] \psi_\mu&(\rep{2},\rep{3}; \rep{4,1})+(\rep{3},\rep{2}; \rep{1,2})\\[3pt]  A_\mu&(\rep{2},\rep{2}; \rep{4,2})\\[3pt]  A_{\mu\nu}&(\rep{3},\rep{1}; \rep{1, 1})+(\rep{1},\rep{3}; \rep{5, 1})\\[3pt] \lambda&(\rep{2,1}; \rep{4,1})+(\rep{1,2}; \rep{5,2})\\[3pt]  \phi &(\rep{1,1}; \rep{5,1})\end{array}
\\
&&\\
\hline
&&\\
\begin{array}{cc}\N=(0,1)&\varnothing \oplus\mathfrak{sp}(1) \\ [3pt] A_\mu&(\rep{2},\rep{2}; \rep{1})\\[3pt] \lambda&(\rep{2,1}; \rep{2})\\
\end{array}
&
\begin{array}{cc}\N=(1,2)&\mathfrak{sp}(1)\oplus\mathfrak{sp}(2) \\ [3pt] g_{\mu\nu}&(\rep{3},\rep{3}; \rep{1,1})\\[3pt] \psi_\mu&(\rep{2},\rep{3}; \rep{2, 1})+(\rep{3},\rep{2}; \rep{1,4})\\[3pt]  A_\mu&(\rep{2},\rep{2}; \rep{2,4})\\[3pt] A_{\mu\nu}&(\rep{3},\rep{1}; \rep{1, 5})+(\rep{1},\rep{3}; \rep{1, 1})\\[3pt] \lambda&(\rep{2,1}; \rep{2,5})+(\rep{1,2}; \rep{1,4})\\[3pt]  \phi &(\rep{1,1}; \rep{1,5})\end{array}
&
\begin{array}{cc}\N=(1,1)&\mathfrak{sp}(1)\oplus\mathfrak{sp}(1) \\ [3pt] g_{\mu\nu}&(\rep{3},\rep{3}; \rep{1,1})\\[3pt] \psi_\mu&(\rep{2},\rep{3}; \rep{2,1})+(\rep{3},\rep{2}; \rep{1,2})\\[3pt] A_\mu&(\rep{2},\rep{2}; \rep{2,2})\\[3pt] A_{\mu\nu}&(\rep{3},\rep{1}; \rep{1, 1})+(\rep{1},\rep{3}; \rep{1, 1})\\[3pt]  \lambda&(\rep{2,1}; \rep{2,1})+(\rep{1,2}; \rep{1,2})\\[3pt]  \phi &(\rep{1,1}; \rep{1,1})\end{array}
 \\
&&\\
\hline
\hline
&&\\
 D=7, \mathfrak{sp}(2)_{ST}
 &
 \begin{array}{cccccc}\multicolumn{3}{c}{\N=1\;\; \mathfrak{sp}(1) }\\ [3pt] A_\mu \; (\rep{5}; \rep{1})& \lambda\;(\rep{4}; \rep{2})&  \phi \;(\rep{1}; \rep{3})\end{array}
  \\
&&\\
\hline
\hline
&&\\
 \begin{array}{cc}\N=1& \mathfrak{sp}(1) \\ [3pt] A_\mu&(\rep{5}; \rep{1})\\[3pt] \lambda&(\rep{4}; \rep{2})\\[3pt]  \phi &(\rep{1}; \rep{3})\end{array}
&\begin{array}{cc}\N=2& \mathfrak{sp}(2) \\ [3pt]g_{\mu\nu}&(\rep{14}; \rep{1})\\[3pt]\psi_{\mu}&(\rep{16}; \rep{4})\\ [3pt]A_\mu, A_{\mu\nu}&(\rep{5}; \rep{10})+(\rep{10}; \rep{5})\\[3pt] \lambda&(\rep{4}; \rep{16})\\  [3pt]\phi &(\rep{1}; \rep{14})\end{array}    \\
&&\\
\hline
\hline
&&\\
   D=8, \mathfrak{su}(4)_{ST}
 &
  \begin{array}{cccc}\multicolumn{3}{c}{\N=1\;\; \mathfrak{u}(1)} \\ [3pt] A_\mu\;(\rep{6}; 0)& \lambda\;(\rep{4}; -1)+(\rep{\overline{4}}; 1)& \phi \;(\rep{1}; 2)+(\rep{1}; -2)\end{array}
  \\
&&\\
\hline
\hline
&&\\
 \begin{array}{cc}\N=1& \mathfrak{u}(1) \\ [3pt] A_\mu&(\rep{6}; 0)\\[3pt] \lambda&(\rep{4}; -1)+(\rep{\overline{4}}; 1)\\[3pt]  \phi &(\rep{1}; 2)+(\rep{1}; -2)\end{array}
&
\begin{array}{cc}\N=2& \mathfrak{u}(2) \\ [3pt]g_{\mu\nu}&(\rep{20}; \rep{1}(0))\\[3pt]\psi_{\mu}&(\rep{20'}; \rep{2}(-1))+(\rep{\overline{20}'}; \rep{2}(1))\\ [3pt]A_\mu&(\rep{6}; \rep{3}(2)+\rep{3}(-2))\\[3pt] A_{\mu\nu}&(\rep{15}; \rep{3}(0))\\[3pt] A_{\mu\nu\rho}&(\rep{10}; \rep{1}(-2))+(\rep{\overline{10}}; \rep{1}(2))\\[3pt]\lambda&(\rep{4}; \rep{2}(-3)+\rep{4}(1))+(\rep{\overline{4}}; \rep{2}(-3)+\rep{4}(1))\\  [3pt]\phi &(\rep{1}; \rep{1}(4)+ \rep{1}(-4)+ \rep{5}(0))\end{array}    \\
&&\\
\hline
\hline
&&\\
   D=9, \mathfrak{so}(7)_{ST}
 &
\N=1\;\; \varnothing \quad A_\mu \; \rep{7}\;\; \lambda\;\rep{8}\;\;  \phi \;\rep{1}
  \\
&&\\
\hline
\hline
&&\\
 \begin{array}{cc}\N=1& \varnothing \\ [3pt] A_\mu&\rep{7}\\[3pt] \lambda&\rep{8}\\[3pt]  \phi &\rep{1}\end{array}
&
\begin{array}{cc}\N=2& \mathfrak{so}(2) \\ [3pt]g_{\mu\nu}&(\rep{27}; 0)\\[3pt]\psi_{\mu}&(\rep{48}; (1)+(-1))\\ [3pt]A_\mu&(\rep{7}; (2)+(0)+(-2))\\[3pt] A_{\mu\nu}&(\rep{21}; (2)+(-2))\\  [3pt]A_{\mu\nu\rho}&(\rep{35}; 0)\\  [3pt]\lambda&(\rep{8}; (3)+(1)+(-1)+(-3))\\  [3pt]\phi &(\rep{1}; (4)+(0)+(-4))\end{array}    \\
&&\\
\hline
\hline
&&\\
   D=10, \mathfrak{so}(8)_{ST}
 &
 \N=(1,0)\;\; \varnothing \quad A_\mu \; \rep{8}_v \;\; \lambda\;\rep{8}_s\\
&&\\
\hline
\hline
&&\\
 \begin{array}{cc}\N=(0,1)& \varnothing \\ [3pt] A_\mu&\rep{8}_v\\[3pt] \lambda&\rep{8}_c\end{array}
&
\begin{array}{cc}\N=(1,1)&\varnothing \\ [3pt]g_{\mu\nu}&\rep{35}_v\\[3pt]\psi_{\mu}&\rep{56}_s+\rep{56}_c\\ [3pt]A_\mu, A_{\mu\nu}, A_{\mu\nu\rho}&\rep{8}_v+\rep{28}_v+\rep{56}_v\\[3pt] \lambda&\rep{8}_s+\rep{8}_c\\  [3pt]\phi & \rep{1}\end{array}    \\
&&\\
\hline
\hline
\end{tabular}
\caption{\footnotesize Tensor products of left and right super Yang-Mills multiplets in $D=6,7,8,9,10$. Dimensions $D=4,5$ are given in \autoref{tab:tensors1}.} \label{tab:tensors2}
\end{table}
\FloatBarrier
On gathering the  spacetime little group representations in \eqref{eq:D5ex}, the $\mathfrak{int}(2,5)\oplus\mathfrak{int}(2,5)=\mathfrak{sp}(2)\oplus\mathfrak{sp}(2)$ representations they carry may be combined into irreducible $\mathfrak{h}(4, 5)=\mathfrak{sp}(4)$ representations,  as illustrated by their  decomposition under $\mathfrak{sp}(4)\supset \mathfrak{sp}(2)\oplus\mathfrak{sp}(2)$:
\be
\begin{split}
(\rep{5}; \rep{1})&\rightarrow (\rep{5}; \rep{1,1}),\\
(\rep{4}; \rep{8})&\rightarrow (\rep{4}; \rep{4,1})+(\rep{4}; \rep{1,4}),\\
(\rep{3}; \rep{27})&\rightarrow (\rep{3}; \rep{1,1})+(\rep{3}; \rep{5,1})+(\rep{3}; \rep{1,5})+ (\rep{3}; \rep{4,4}),\\
(\rep{2}; \rep{48})&\rightarrow (\rep{2}; \rep{4,1})+(\rep{2}; \rep{1,4})+(\rep{2}; \rep{4,5})+(\rep{2}; \rep{5,4}),\\
(\rep{1}; \rep{42})&\rightarrow (\rep{1}; \rep{1,1})+ (\rep{1}; \rep{4,4})+(\rep{1}; \rep{5,5}).
\end{split}
\ee

\subsection{R-symmetry algebras}\label{Toy Model: obtaining the R-symmetry algebras of supergravity from super Yang-Mills}

 We begin with the simple relationship between the R-symmetry algebras of  supergravity  and  its generating  super Yang-Mills factors. While somewhat trivial this example introduces much of the notation and concepts needed later for the $H$ and $G$  algebras.  
 
 R-symmetry  is defined here as the automorphism group of the supersymmetry algebra. Its action  on the $\N$-extended supersymmetry generators $Q$ is given schematically by
\be
[T_A,Q_{a}]=(U_A)_a{}^b Q_{b}
,\qquad a, b =1,\ldots,\N.\ee
 The R-symmetry algebra is fixed by the reality properties of the minimal spinor representation in $D$ mod 8 dimensions. See, for example, \cite{Strathdee:1986jr}. 
 
 %In summary,
 %\be\label{eq:Rsym}
%\mathfrak{a}(\mathcal{N}_L+\N_R, \mathds{D})=\mathfrak{a}(\mathcal{N}_L, \mathds{D})\oplus\mathfrak{a}(\mathcal{N}_R,\mathds{D})+\mathds{D}[\mathcal{N}_L, \mathcal{N}_R]
%\ee 
  %gives the  R-symmetry algebra, $\mathfrak{ra}(\mathcal{N}, D)=\mathfrak{a}(\mathcal{N}, \mathds{D})$,  of the supergravity  in terms of the R-symmetry algebras of the left/right super Yang-Mills factors.   

Making use of the super-Jacobi identities, it can be shown that the $U_A$'s form a representation of the algebra $\mathfrak{so}(\mathcal{N}), \mathfrak{u}(\mathcal{N}), \mathfrak{sp}(\mathcal{N})$ for $Q$ real, complex,  quaternionic (pseudoreal), respectively.  For $s-t=2,3,\ldots10 \mod 8$, where $D=t+s$, the spinor representations follow the famous Bott periodic sequence $\C, \Q, \Q\oplus\Q, \Q, \C, \R, \R\oplus\R, \R, \C, \Q, \Q\oplus\Q, \Q,\ldots$ Since R-symmetry  commutes with the Lorentz algebra, only the reality properties of the spinor representation and $\N$ are relevant. Consequently, we may associate a (direct sum of) division algebra(s), denoted $\mathds{D}$, to every dimension, as given in \autoref{tab:AD}, which will then dictate  the R-symmetry algebra. The identification of $\mathds{D}$ for each $D=3,\ldots,10$ follows from the close relationship between Clifford and division algebras. 
 \newcolumntype{M}{>{$}c<{$}}
\begin{table}[h]
\centering
\begin{tabular}{M|M|M|M|M }
 \hline
 \hline
D&\Cliff(D-3)\cong\Cliff_0(D-2) & \mathds{D} &\begin{array}{l} D-2 ~\text{spinor representation}\\\cong D-3 ~ \text{pinor representation}\end{array}&\text{R-symmetry algebra}  \\
\hline
10&\R[8]\oplus \R[8]  &\R^{+}\oplus \R^{-}&\R^{8}_+\oplus\R^{8}_{-}&\mathfrak{so}(\mathcal{N}_+) \oplus\mathfrak{so}(\mathcal{N}_-)  \\
9&\R[8]& \R &\R^8 &\mathfrak{so}(\mathcal{N}) \\
8&\C[4]&\C & \C^4&\mathfrak{u}(\mathcal{N})\\
7&\Q[2]&\Q &\Q^2&\mathfrak{sp}(\mathcal{N}) \\
6 &\Q[1]\oplus \Q[1] &\Q^{+}\oplus \Q^{-}  & \Q_+\oplus \Q_- &\mathfrak{sp}(\mathcal{N}_+) \oplus \mathfrak{sp}(\mathcal{N}_-)  \\
5 &\Q[1]&\Q & \Q &\mathfrak{sp}(\mathcal{N})  \\
4 &\C[1]&\C &\C &\mathfrak{u}(\mathcal{N})   \\
3 &\R[1]&\R&  \R &\mathfrak{so}(\mathcal{N})   \\
 \hline
 \hline
\end{tabular}
\caption{\footnotesize The Clifford (sub)algebras,  $\mathds{D}$,  spinor representation and R-symmetry algebra for dimensions $D=3,\ldots,10$.} \label{tab:AD}
\end{table}

Let us briefly review these ideas here. For a detailed survey  see \cite{harvey1990spinors, Baez:2001dm} and the references therein.  
 For a unital algebra $A$ let $A[m, n]$ denote  the set of $m\times n$ matrices with entries in $A$. When $m=n$ we will also write $A[n]$. For $D=3,\ldots, 10$ the Euclidean Clifford algebra $\Cliff(D-3)$ can be mapped to the (direct sum of) matrix algebras  $\alg[n]$, as given in \autoref{tab:AD}. Up to equivalence, the unique non-trivial irreducible representations  of  $\alg[n]$ and  $\alg[n]\oplus \alg[n]$  are   $\alg^n$ and  $\alg^n\oplus\alg^n$, respectively.  These  representations restrict to the  \emph{pinors}    of  $\Pin(D-3)$, the double cover of  $\Orth(D-3)$, as it is generated by the  subset  of unit vectors in $\R^{D-3}$. There is a canonical isomorphism from $\Cliff(D-3)$ to $\Cliff_0(D-2)$, where $\Cliff_0(m)$ denotes the subalgebra generated by products of an even number of vectors in $\R^{m}$. Since $\Spin(D-2)$, the double cover of the spacetime little group, sits inside $\Cliff_0(D-2)$ as the set of all elements that are a product  of unit vectors, the pinors of $\Pin(D-3)$ are precisely the spinors of $\Spin(D-2)$, as given in \autoref{tab:AD}. The supersymmetry algebra generators, $Q$, transform according as these representations under $\Spin(D-2)$.  Hence, we may identify $\mathds{D}$ as the appropriate algebra for each spacetime dimension $D$. Note that in dimensions 6 and 10 the direct sum structure of $\mathds{D}$ corresponds to  the existence of  $\N=(\N_+, \N_-)$ chiral theories.

Let us now briefly recall some of the standard relations between $\R, \C, \Q$ and the classical Lie algebras. Denote by $\mathfrak{a}(n,\alg)$   the set of  anti-Hermitian elements  in $\mathds{A}[n]$,
\be
\label{eq:antihermA}
\mathfrak{a}(n,\alg):=\{x\in \alg[n]:x^\dagger=-x \}.
\ee
Using the standard matrix commutator these constitute the classical Lie algebras
\be
\mathfrak{a}(n,\alg)=\left\{\begin{array}{lll}
\mathfrak{so}(n),&&\alg=\R;\\ 
\mathfrak{u}(1)\oplus\mathfrak{su}(n), &&\alg=\C;\\ 
\mathfrak{sp}(n),&&\alg=\Q.
\end{array}\right.
\ee
Let $\mathfrak{sa}(n,\alg)$ denote their special subalgebras:
\be
\begin{array}{lllllllll}
\mathfrak{sa}(n,\R)&:=&\{x\in \alg[n]:x^\dagger&=-x \}&=&\mathfrak{so}(n);\\
\mathfrak{sa}(n,\C)&:=&\{x\in \alg[n]:x^\dagger&=-x, \tr(x)=0\}&=&\mathfrak{su}(n);\\
\mathfrak{sa}(n,\Q)&:=&\{x\in \alg[n]:x^\dagger&=-x \}&=&\mathfrak{sp}(n).
\end{array}
\ee
The seemingly undemocratic definition of $\mathfrak{sa}(n,\alg)$  follows naturally from the geometry of  projective spaces since
\be
\mathfrak{Isom}(\alg \mathds{P}^{n-1})\cong \mathfrak{sa}(n,\alg)
\ee
for $\alg=\R, \C, \Q$. 
In the octonionic case only $\Oct\mathds{P}^{1}$ and  $\Oct\mathds{P}^{2}$ constitute projective spaces with $\mathfrak{Isom}(\Oct\mathds{P}^{1})\cong\mathfrak{so}(8)$ and $\mathfrak{Isom}(\Oct\mathds{P}^{2})\cong \mathfrak{f}_{4(-52)}$, reflecting their exceptional status.

 It then follows that the  $\N$-extended R-symmetry algebras  in $D$ dimensions, denoted $\mathfrak{ra}(\N, D)$, are  given by 
 \be
\mathfrak{ra}(\N, D)=\mathfrak{a}(\N, \mathds{D}),
\ee
where for $\N=(\N^+, \N^{-})$, as is the case for $D=6, 10$, we have used the definition 
\be
\mathds{D}[(\N^+, \N^{-})]:=\mathds{D}^{+}[\N^{+}]\oplus \mathds{D}^{-}[\N^{-}].
\ee

 Since $\mathds{D}[m, n]\cong \mathds{D}^{m}\otimes\mathds{D}^{n}$ forms a  natural (but not necessarily irreducible) representation of $\mathfrak{a}(m, \mathds{D})\oplus\mathfrak{a}(n,\mathds{D})$, it follows quite simply that the $(\mathcal{N}_L+\mathcal{N}_R)$-extended R-symmetry algebra is given by the $\mathcal{N}_L$ and $\mathcal{N}_R$ R-symmetry algebras via
 \be\label{eq:raformula2}
\mathfrak{ra}(\mathcal{N}_L+\mathcal{N}_R,D)=\mathfrak{a}(\mathcal{N}_L+\mathcal{N}_R, \mathds{D})=\mathfrak{a}(\mathcal{N}_L, \mathds{D})\oplus \mathfrak{a}(\mathcal{N}_R, \mathds{D})+\mathds{D}[\mathcal{N}_L, \mathcal{N}_R].
\ee
The commutators for elements $X_L\in\mathfrak{a}(\mathcal{N}_L, \mathds{D})$, $X_R\in\mathfrak{a}(\mathcal{N}_R, \mathds{D})$ and $M, N\in\mathds{D}[\mathcal{N}_L, \mathcal{N}_R]$ are given by
\be\label{eq:racomms}
\begin{array}{llllllllll}
&[X_L, M]&=& X_LM &\in \mathds{D}[\mathcal{N}_L, \mathcal{N}_R],\\ [8pt]
&[X_R, M]&=& -MX_R &\in \mathds{D}[\mathcal{N}_L, \mathcal{N}_R],\\[8pt]
&[M, N]&=& (NM^\dagger - MN^\dagger)\oplus (N^\dagger M - M^\dagger N) &\in  \mathfrak{a}(\mathcal{N}_L, \mathds{D})\oplus \mathfrak{a}(\mathcal{N}_R, \mathds{D}).
\end{array}
\ee
These commutation relations follow from the standard matrix commutators of 
\be\label{eq:radecomp}
X= \begin{pmatrix}X_L&0\\0&X_R \end{pmatrix}+ \begin{pmatrix}0&M\\-M^\dagger &0\end{pmatrix},
\ee
where $X\in \mathfrak{a}(\N_L+\N_R, \mathds{D})$. Note, as a  $\mathfrak{a}(\mathcal{N}_L, \C)\oplus \mathfrak{a}(\mathcal{N}_R, \C)$-module $\C[\mathcal{N}_L, \mathcal{N}_R]$ is not   irreducible. For example, in the maximal $D=4$ case it 
 corresponds to the $(\rep{4},\rep{\overline{4}})+(\rep{\overline{4}}, \rep{4})$ representation of $\mathfrak{su}(4)\oplus \mathfrak{su}(4)\cong\mathfrak{sa}(4, \C)\oplus \mathfrak{sa}(4, \C)$. The formula \eqref{eq:raformula2} and its commutators \eqref{eq:racomms} amount to the well-known statement that the pairs $[\mathfrak{so}(p+q), \mathfrak{so}(p)\oplus\mathfrak{so}(q)]$, $[\mathfrak{su}(p+q), \mathfrak{su}(p)\oplus\mathfrak{su}(q)\oplus\mathfrak{u}(1)]$ and $[\mathfrak{sp}(p+q), \mathfrak{sp}(p)\oplus\mathfrak{sp}(q)]$ constitute type I symmetric spaces. 

From the perspective of the left/right tensor product,  $\mathfrak{a}(\mathcal{N}_L, \mathds{D})\oplus \mathfrak{a}(\mathcal{N}_R, \mathds{D})$ is generated directly by the R-symmetries of the left and right factors acting on $Q_L$ and $Q_R$ independently. However,   together they form an irreducible doublet $(Q_L, Q_R)\in\mathds{D}^{\N_L}\oplus\mathds{D}^{\N_R}$ (suppressing the spacetime little group spinor representation space), which must be rotated by an $\mathfrak{a}(\mathcal{N}_L, \mathds{D})\oplus \mathfrak{a}(\mathcal{N}_R, \mathds{D})$-module. The most general consistent subset of $\End(\mathds{D}^{\N_L}\oplus\mathds{D}^{\N_R})$ is given by $\mathds{D}[\mathcal{N}_L, \mathcal{N}_R]$, which completes \eqref{eq:raformula2} as is made clear by \eqref{eq:radecomp}. In the sense to be described in \autoref{sec:Hsym}, these additional elements can be  generated by  $Q_L\otimes Q_R\in\mathds{D}[\mathcal{N}_L, \mathcal{N}_R]$ by formally neglecting its little group representation space.

It follows from \cite{Borsten:2013bp} that for $D=3$ the R-symmetry algebras admit an alternative  Freudenthal magic square description. Recall that the U-duality groups in $D=3$  form the Freudenthal Magic square given by
\be
\begin{split}\label{magicSquareJordan}
\mathfrak{L}_3(\alg_L,\alg_R)&=\mf{tri}(\alg_{ \mathcal{N}_L})\oplus \mf{tri}(\alg_{ \mathcal{N}_R})+3(\alg_{ \mathcal{N}_L}\otimes\alg_{ \mathcal{N}_R})\\
&=\mathfrak{der}\alg_{\N_L}\oplus\mathfrak{der}\mathfrak{J}_{3}(\alg_{\N_R})+\text{Im}\alg_{\N_L}\otimes\mathfrak{J}^{0}_{3}(\alg_{\N_R}),
\end{split}
\ee  
where $\mathfrak{der}$ denotes the derivation algebra,  $\mathfrak{J}_3(\alg)$ is the Jordan algebra of $3\times 3$ Hermitian matrices over $\alg$ and $\mathfrak{J}^{0}_{3}(\alg)$ is its traceless subspace. See for example \cite{Baez:2001dm}. One can generalise this construction for any rank of the Jordan algebra $\mathfrak{J}_n(\alg)$,
\be
\mathfrak{L}_n(\alg_{\N_L},\alg_{\N_R})
=\mathfrak{der}\alg_{\N_L}\oplus\mathfrak{der}\mathfrak{J}_n(\alg_{\N_R})+\text{Im}\alg_{\N_L}\otimes\mathfrak{J}^{0}_{n}(\alg_{\N_R}),
\ee  
where for $n>3$ we must exclude the octonionic case \cite{Barton:2003}. The supergravity R-symmetry algebras   in $D=3$ are  given by  $n=2$, 
\be
\mathfrak{L}_2(\alg_{\N_L}, \alg_{\N_R})=\mathfrak{der}\alg_{\N_L}\oplus\mathfrak{der}\mathfrak{J}_2(\alg_{\N_R})+\text{Im}\alg_{\N_L}\otimes\mathfrak{J}^{0}_{2}(\alg_{\N_R})=\mathfrak{so}(\mathcal{N}_L+\mathcal{N}_R).
\ee
%\FloatBarrier
 %\begin{table}[ht]
 %\begin{center}
%\begin{ruledtabular}
%\begin{tabular}{c|ccccccc}
%\hline
%\hline
 %$\alg_L/\alg_R$ && $\R$ & $\C$  & $\Q$  & $\Oct$ & \\
 %\hline
 %\\
   %$\R$ && $\mathfrak{so}(2)$ & $\mathfrak{so}(3)$   & $\mathfrak{so}(5)$   & $\mathfrak{so}(9)$  & \\
  %$\C$ && $\mathfrak{so}(3)$ & $\mathfrak{so}(4)$   & $\mathfrak{so}(6)$   & $\mathfrak{so}(10)$   &\\
  %$\Q$ && $\mathfrak{so}(5)$ & $\mathfrak{so}(6)$   & $\mathfrak{so}(8)$   & $\mathfrak{so}(12)$  & \\
   %$\Oct$ && $\mathfrak{so}(9)$ & $\mathfrak{so}(10)$   & $\mathfrak{so}(12)$   & $\mathfrak{so}(16)$&   \\
   %\\
   %\hline
   %\hline
%\end{tabular}
%\caption{The magic square $\mathcal{M}_2(\alg_L,\alg_R)$ corresponds to the R-symmetry of supergravities obtained by tensoring $\mathcal{N}_L=|\alg_L|$ and $\mathcal{N}_R=|\alg_R|$ super Yang-Mills theories in 3 dimensions.}
%\end{ruledtabular}
 %\end{center}
%\end{table}

\subsection{$H$ algebras}\label{sec:Hsym}

  With this construction   in  mind we  turn our attention now to the algebra $\mathfrak{h}$ of the maximal compact subgroup $H\subset G$ and, in particular, how it is built from the global symmetries of the left and right super Yang-Mills theories.

We will write $\mathfrak{h}(\N_L+\N_R, D)$ in terms of $\mathfrak{int}(\N_L, D)$ and $\mathfrak{int}(\N_R, D)$.  First, note that $\mathfrak{int}$ and $\mathfrak{h}$ have a similar structure; they are both given by $\mathfrak{a}(\N, \mathds{D})$, possibly with  additional commuting factors, which we denote by $\mathfrak{p}$. From  \autoref{R-symmetries in all dimensions table} we observe that almost uniformly,
\be\label{rule}
\mathfrak{int}(\N, D)=\mathfrak{a}(\mathcal{N}, \mathds{D})\oplus\mathfrak{p},
\ee
where $\mathfrak{p}= \mathfrak{so}(2), \mathfrak{so}(3)$ for  $D=3, \N=2, 4$. The extra factors in $D=3$  follow from dualising the gauge field into a scalar, which  also enhances  $\mathfrak{so}(7)\rightarrow\mathfrak{so}(8)$ in the maximally supersymmetric case. These symmetries are given by the triality algebra \cite{Borsten:2013bp},
\be
\mathfrak{int}(\N, 3)=\mathfrak{tri}(\alg_\N), \qquad\alg_\N=\R, \C, \Q, \Oct. 
\ee 
Here the fields of the $\mathcal{N}=1,2,4,8$ super Yang-Mills belong to $\alg_\N=\R,\C,\Q,\Oct$. Then $\mathfrak{p}=\mathfrak{so}(2)=\mathfrak{u}(1)$ is generated by $i$ and $\mathfrak{so}(3)=\mathfrak{sp}(1)$ is generated by $i,j,k$, the imaginary unit quaternions acting from the right, which commutes with the usual left actions.
These commuting factors of the  left and right super Yang-Mills theories are inherited in  the tensor product and, hence, the resulting $\mathfrak{h}$ also contains a commuting $\mathfrak{p}_L\oplus\mathfrak{p}_R$. 
The $\mathfrak{u}(1)$ factors  in $D=4$ correspond to having to include CPT conjugates.   The  exception to \eqref{rule} is the maximally supersymmetric $D=4$ case, which is reduced to $\mathfrak{sa}(\mathcal{N}, \mathds{D})$ as no CPT conjugate need be added. 
\newcolumntype{M}{>{$}l<{$}}
\newcolumntype{C}{>{$}c<{$}}
\begin{table}[h]
\centering
\begin{tabular}{C|CC|CC|CC|CCC}
\hline
\hline
 D &\multicolumn{2}{C|}{Q=16}  &\multicolumn{2}{C|}{Q=8}  &\multicolumn{2}{C|}{Q=4}&\multicolumn{2}{C}{Q=2}\\
&\N&\mathfrak{int}&\N&\mathfrak{int}&\N&\mathfrak{int}&\N&\mathfrak{int}& \\
\hline
10 &1&\varnothing   &- &  &- & &-  \\
9 &1&   \varnothing  &-  & &- &&-  \\
8 &1 &  \mathfrak{u}(1) &- &  &-& &- \\
7 &1&  \mathfrak{sp}(1)    &-&  &- &&- \\
\multirow{2}{*}{6} & (1,1)&  \mathfrak{sp}(1)\oplus\mathfrak{sp}(1)    &(1,0)& \mathfrak{sp}(1)\oplus\varnothing &-&&- \\
 & (2,0)\ & \mathfrak{sp}(2)\oplus\varnothing    &(1,0)& \mathfrak{sp}(1)\oplus\varnothing  &-&&-  \\
5 & 2 & \mathfrak{sp}(2)    &1& \mathfrak{sp}(1)  &- &&- \\
4 & 4 & \mathfrak{su}(4)    &2&  \mathfrak{su}(2)\oplus \mathfrak{u}(1)  &1 &  \mathfrak{u}(1) &- \\
3\ & 8 & \mathfrak{so}(7)    &4&  \mathfrak{so}(4)  &2 & \mathfrak{so}(2) & 1& \varnothing \\
3^{*}\ & 8 & \mathfrak{so}(8)    &4&  \mathfrak{so}(4)\oplus \mathfrak{so}(3)  &2 & \mathfrak{so}(2)\oplus \mathfrak{so}(2) & 1& \varnothing \\
\hline
\hline
\end{tabular}\caption{\small The internal global symmetry algebras  $\mathfrak{int}(\N, D)$ of super Yang-Mills theories in $D\geq 3$. In $D=6$ we have  included the $(2, 0)$ and $(1, 0)$ tensor multiplets. Note, for  $D=3^{*}$ we have  dualised the vector yielding an enhanced  symmetry, $\mathfrak{int}(\N, 3)=\mathfrak{tri}(\alg_\N)$ for $\alg_\N=\R, \C, \Q, \Oct$. For interacting Lagrangians this symmetry is reduced to the intermediate algebra $\mathfrak{int}(\alg_\N):=\{(A, B , C)\in\mathfrak{tri}(\alg_\N) | A(1)=0\}$, which  gives $\varnothing, \mathfrak{so}(2), \mathfrak{so}(4), \mathfrak{so}(7)$ for $\alg_\N=\R, \C, \Q, \Oct$, respectively.  The enhanced $\mathfrak{tri}(\alg_\N)$ symmetry is recovered in the infrared limit. Note, while in general the R-symmetry algebra $\mathfrak{ra}(\N, D)$ and  the internal global symmetry algebra  $\mathfrak{int}(\N, D)$ coincide,  there are several exceptions such as $\mathfrak{su}(4)$ versus $\mathfrak{u}(4)$ for $D=4, \N=4$.}\label{R-symmetries in all dimensions table}
\end{table}
   \begin{figure}
\begin{center}
\includegraphics[scale=0.425]{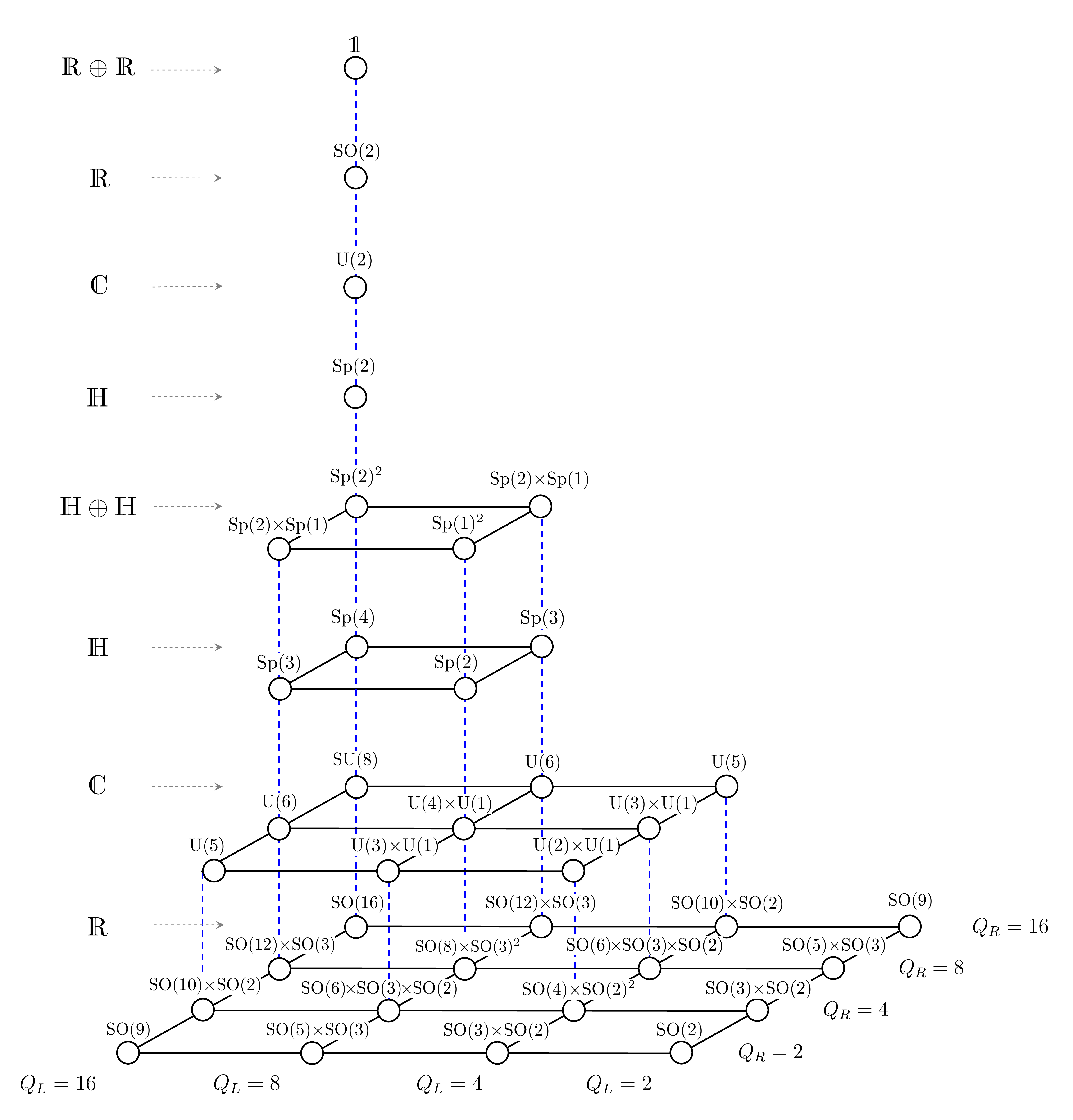}
\caption{\footnotesize Pyramid of maximal compact subgroups $H\subset G$. The amount of supersymmetry is determined by the horizontal axes. The spacetime dimension is determined by the division algebra $\mathds{D}$ on the vertical axis as given in \autoref{tab:AD}.}\label{fig:Halldims} 
\end{center}
\end{figure}

The algebras $\mathfrak{h}(\N_L+\N_R, D)$ presented in \autoref{tab:tensors1} and \autoref{tab:tensors2} are consequently  given by
 \be\label{eq:Halg2}
 \begin{split}
\mathfrak{h}(\mathcal{N}_L+\mathcal{N}_R,D)&=\mathfrak{int}(\mathcal{N}_L,D)\oplus\mathfrak{int}(\mathcal{N}_R,D)\oplus\delta_{D, 4}\mathfrak{u}(1)+\mathds{D}[\mathcal{N}_L, \mathcal{N}_R]\\
&=\mathfrak{sa}(\mathcal{N}_L+\mathcal{N}_R, \mathds{D})\oplus\mathfrak{p}_L\oplus\mathfrak{p}_R,\\
\end{split}
\ee 
where the non-trivial commutators are those given in \eqref{eq:racomms}.  In $D=3$ $\mf{p}_{L/R}$ are given by $\varnothing, \mf{so}(2)$ or $\mf{so}(3)$. In $D=4$ $\mf{p}_{L/R}$ are given by $\varnothing$ or $\mf{u}(1)$. In $D=8$ there is a single  $\mf{u}(1)$ factor. Special care needs to be taken for the  $\mathfrak{u}(1)$ factors in $D=4$; a linear combination of their generators is taken such that the $\mathds{D}[\mathcal{N}_L, \mathcal{N}_R]$ generators have charge $\pm2$ under a single $\mf{u}(1)$, according as the Kantor-Koecher-Tits 3-grading
\be
\mf{su}(m+n)=(\rep{m},\overline{\rep{n}})_{\sst{(-2)}}+[\mf{u}(1)\oplus\mf{su}(m)\oplus\mf{su}(n)]_{\0}+(\overline{\rep{m}},{\rep{n}})_{\sst{(2)}},
\ee
 and are uncharged under the remaining factors. The corresponding pyramid of $H$ groups, which generalises the magic $H$ pyramid of \cite{Anastasiou:2013hba},  is presented in \autoref{fig:Halldims}.
 
 Clearly, the term 
$\mathfrak{int}(\mathcal{N}_L,D)\oplus\mathfrak{int}(\mathcal{N}_R,D)
$ 
follows directly from the left and right super Yang-Mills symmetries. It acts on the gravitini $\psi_{L}$ and $\psi_{R}$ independently in the defining representation, since the left and right gauge potentials $A_{L/R}$ are $\mathfrak{int}(\mathcal{N}_{L/R},D)$ singlets. However,  as for the supersymmetry charges, the gravitini are collected into an irreducible doublet $(\psi_{L}, \psi_{R})$, which is rotated by $\mathds{D}[\mathcal{N}_L, \mathcal{N}_R]$, and hence  transform in the defining representation of 
\be
\mathfrak{sa}(\mathcal{N}_L+\mathcal{N}_R, \mathds{D})=\mathfrak{sa}(\mathcal{N}_L, \mathds{D})\oplus\mathfrak{sa}(\mathcal{N}_R, \mathds{D})\oplus\delta_{D, 4}\mathfrak{u}(1)+\mathds{D}[\mathcal{N}_L, \mathcal{N}_R].
\ee

While the $\mathds{D}[\mathcal{N}_L, \mathcal{N}_R]$ component of $\mathfrak{h}$ is implied by consistency, one might also more ambitiously ask whether  it can be directly generated by elementary operations acting on the left and right super Yang-Mills fields in same way  $\mathfrak{int}(\mathcal{N}_L,D)\oplus\mathfrak{int}(\mathcal{N}_R,D)\subset\mathfrak{h}$  obviously is. Having already used all  left/right bosonic symmetries, only the  left/right supersymmetry generators remain. The conventional infinitesimal supersymmetry variation of the $left\otimes right$   states correctly gives the infinitesimal supersymmetry variation on the corresponding supergravity states \cite{Siegel:1995px, Bianchi:2008pu, Anastasiou:2014qba}. Seeking, instead,  internal \emph{bosonic} transformations on the supergavity multiplet that follow from supersymmetry  on the left and right Yang-Mills multiplets suggests starting from the rather unconventional tensor product of the left and right supercharges, $Q\otimes \tilde{Q}$. That this might work, at least formally, follows from the observation
\be
Q\in \mathds{D}^{\N}\quad  \Rightarrow \quad Q\otimes \tilde{Q}\in \mathds{D}^{\N_L} \otimes \mathds{D}^{\N_R} \cong\mathds{D}[\N_L, \N_R],
\ee
where we are explicitly suppressing the spacetime indices.

Adopting the spinor-helicity formalism, a simple concrete example sufficient to illustrate the principle is given by the $4+4$ positive helicity gravitini states of $D=4, \N=8$ supergravity, 
\be\label{eq:psistates}
\psi^{a}_{+}=\lambda^{a}_{+}\otimes \tilde{A}_{+}, \qquad \psi^{a'}_{+}=A_{+}\otimes \tilde{\lambda}^{a'}_{+},
\ee
%\be
%\begin{pmatrix}\psi^{a}_{+}\\\psi^{a}_{-}\end{pmatrix}=\begin{pmatrix}\chi^{a}_{+}\otimes \tilde{A}_{+}\\\chi^{a}_{-}\otimes \tilde{A}_{-}\end{pmatrix}, \qquad \begin{pmatrix}\psi^{a'}_{+}\\\psi^{a'}_{-}\end{pmatrix}=\begin{pmatrix}A_{+}\otimes \tilde{\chi}^{a'}_{+}\\A_{-}\otimes \tilde{\chi}^{a'}_{-}\end{pmatrix},
%\ee
where $a, a'=1,\ldots,4$ are the $\rep{4}$ of $\mathfrak{su}(4)_{L}$ and $\mathfrak{su}(4)_R$, respectively. Defining $Q^{a}_{-}=-\epsilon^{\alpha}Q_{\alpha}^{a}$ and $Q^{+}_{a}=-\epsilon_{\dot{\alpha}}Q^{\dot{\alpha}}_{a}$, the relevant super Yang-Mills transformations are
\be
\begin{array}{lll}
Q_{a}^{+}A_{+}(p)=0,& \qquad &Q^{+}_{a}\lambda_{+}^{b}=\langle \epsilon p\rangle \delta^{b}_{a} A_{+}(p),\\
Q^{a}_{-}A_{+}(p)=[p\epsilon]\lambda_{+}^{a}(p),& \qquad &Q_{-}^{a}\lambda_{+}^{b}=[ p \epsilon ]  \phi^{[ab]}(p).
\end{array}
\ee
Applying these to \eqref{eq:psistates} we obtain
 \be
\begin{array}{lllllllll}
[Q_{a}^{+}\otimes \tilde{Q}_{-}^{a'}] \psi^{b}_{+} &= &[Q_{a}^{+}\lambda^{b}_{+}]&\otimes& [\tilde{Q}_{-}^{a'}\tilde{A}]&=&[p\epsilon] \langle \epsilon p\rangle\delta^{b}_{a} A_{+} \otimes \tilde{\lambda}_{+}^{a'}&=&[p\epsilon] \langle \epsilon p\rangle\delta^{b}_{a}  \psi^{a'}_{+},\\[3pt]
[Q_{a}^{+}\otimes \tilde{Q}_{-}^{a'}] \psi^{b'}_{+} &=& [Q_{a}^{+}A_{+}]&\otimes& [\tilde{Q}_{-}^{a'}\tilde{\lambda}^{b'}]&=&0,\\[3pt]
[Q^{a}_{-}\otimes \tilde{Q}^{+}_{a'}] \psi^{b}_{+} &= &[Q^{a}_{-}\lambda^{b}_{+}]&\otimes& [\tilde{Q}^{+}_{a'}\tilde{A}]&=&0,\\[3pt]
[Q^{a}_{-}\otimes \tilde{Q}^{+}_{a'}] \psi^{b'}_{+} &=& [Q^{a}_{-}A_{+}]&\otimes& [\tilde{Q}^{+}_{a'}\tilde{\lambda}^{b'}]&=&[p\epsilon] \langle \epsilon p\rangle\delta^{b'}_{a'}  \lambda_{+}^{a}\otimes \tilde{A}_{+}&=&[p\epsilon] \langle \epsilon p\rangle\delta^{b'}_{a'}  \psi^{a}_{+},
\end{array}
\ee
which, up to the factors of  $[p\epsilon] \langle \epsilon p\rangle$, is precisely the action of the $\mathfrak{su}(8)$ generators belonging to $\mathds{D}[\mathcal{N}_L, \mathcal{N}_R]$  on the positive helicitly gravitini states valued in $\mathds{D}^{\N_L} \oplus \mathds{D}^{\N_R}$, for $D=\N_{L/R}=4$, or,  in a perhaps more familiar language, the action of generators in the $\rep{(4,\overline{4})+(\overline{4}, 4)}$ component of $\mathfrak{su}(8)$ acting on the $\rep{8}=\rep{(4,1)+(1, 4)}$ representation. 

Thus, formally suppressing the spacetime components of the supercharges (and  parameters)  provides a definition of the elementary transformations acting on the left and right states, which correctly reproduces the action of $\mathfrak{h}$ on their tensor product. More concretely, we have
\be\label{oldQs}
Q_-^a=\int \frac{d^3p}{(2\pi)^3 2E_p}[p\epsilon]\left[-\lambda_+^a (A_+)^\dagger+\phi^{[ab]}(\lambda^{b}_{+})^\dagger+2\lambda_{b-}(\phi^{[ab]})^\dagger-A_-(\lambda_{a-})^\dagger\right],
\ee which ensures the correct action of the supersymmetry operator\footnote{See \cite{Bianchi:2008pu} for the full set of supersymmetry transformations.} 
 with non-trivial equal time (anti)commutation relations:
\be
\begin{aligned}
[A_\pm(p),A^\dagger_\pm(q)]&=(2\pi)^32E_p\delta^3(\vec{p}-\vec{q}),\\
\{\lambda_\pm^a(p),\lambda_{b\pm}^\dagger(q)\}&=(2\pi)^32E_p\delta^3(\vec{p}-\vec{q})\delta^a_b,\\
[\phi^{[ab]}(p),\phi_{[cd]}(q)]&=(2\pi)^32E_p\delta^3(\vec{p}-\vec{q})\delta^{[ab]}_{[cd]},
\end{aligned}
\ee
where $\phi^{[ab]}=\frac{1}{4!}\varepsilon^{abcd}\phi_{[cd]}$. The operators $Q_L\otimes Q_R$ generating $\mathds{D}[\mathcal{N}_L, \mathcal{N}_R]$ are then defined  by simply dropping the  $[p\epsilon]$ factors in  this representation of $Q$. For example:
\be
Q_{L}{}^{a}_{\mp}:=\int \frac{d^3p}{(2\pi)^3 2E_p}\left[-\lambda_{\pm}^a(A_{\pm})^\dagger +\phi^{[ab]}(\lambda^{b}_{\pm})^\dagger+2\lambda_{b{\mp}}(\phi^{[ab]})^\dagger-A_{\mp}(\lambda_{a\mp})^\dagger\right]
\ee
and similarly for the remaining $Q$'s. 

One can check this construction gives the correct action on the rest of the $\N=8$ multiplet and   generalises to any dimension and number of supercharges. Note that in higher dimensions, where the little group is larger than $\mathfrak{u}(1)$, the tensor product of two super Yang-Mills states typically yields a direct sum of supergravity states; to pick a specific component we need to project out the desired representation. To find the action of $ Q_L\otimes Q_R$ on a state, we first act on the tensor product which contains it, and then project out the state we want. Returning to our maximal $D=5$ example (described in \eqref{eq:D5ex} and \autoref{tab:tensors1}), we see that the gravitini states live in the $(\rep{4}; \rep{1,4})+(\rep{4}; \rep{4,1})$ representation of $\mathfrak{so}(3)_{ST}\oplus\mathfrak{sp}(2)\oplus\mathfrak{sp}(2)$. Focusing on the $(\rep{4}; \rep{1,4})$ states, we see that they are obtained by a projection of $A_\mu\otimes\tilde{\lambda}=(\rep{3}; \rep{1})\otimes(\rep{2}; \rep{4})=(\rep{4}; \rep{1,4})+(\rep{2}; \rep{1,4})$. Then action of  $Q_L$ and $Q_R$ , both living in $(\rep{2}; \rep{4})$ of $\mathfrak{so}(3)_{ST}\oplus\mathfrak{sp}(2)$, gives $(\rep{4}; \rep{4,1})+(\rep{2}; \rep{4,1})+(\rep{2}; \rep{4,5})$. Projecting out the gravitini, we find that the $(\rep{4}; \rep{1,4})$ states have been rotated into $(\rep{4}; \rep{4,1})$ states.

\subsection{$G$ algebras} \label{sec:Gsym}

   \begin{figure} 
\hspace{-0.25in}\includegraphics[scale=0.425]{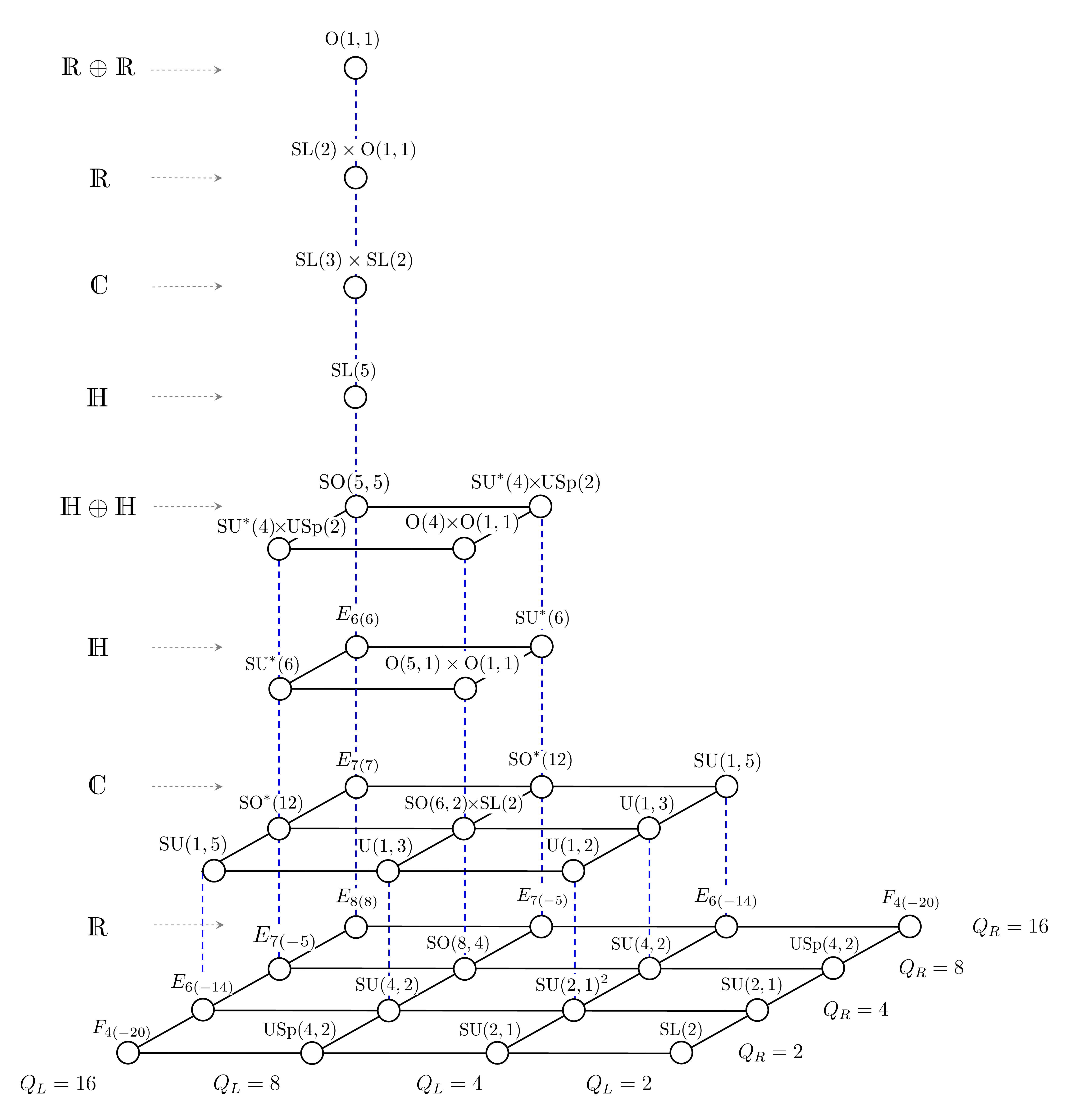}
\begin{center}
\caption{The U-duality group $G$ in all dimensions.}\label{fig:Galldims}
\end{center}
\end{figure}

 The non-compact U-duality algebras of the supergravity theories appearing  in the pyramid,  \autoref{fig:Galldims}, can be built straightforwardly using the  tensor product of the left and right super Yang-Mills multiplets.  Recall,   the scalars of supergravity coupled to matter generated by squaring parametrise a $G/H$ coset and $T_{p}(G/H)\cong\mathfrak{p}=\mathfrak{g}\ominus\mathfrak{h}$, and so carry the $\mathfrak{p}$-representation of $H$. Consequently, the non-compact generators $\mathfrak{p}$, in a manifest $\mathfrak{int}(\N_L, D)\oplus\mathfrak{int}(\N_R, D)$ basis, can be read off from those tensor products which yield scalars, which are schematically  given by:
\be\label{eq:scalars}
A_{\mu}\otimes\tilde{A}_{\nu}, \qquad\lambda^{a}\otimes \tilde{\lambda}^{a'}, \qquad \phi^{i}\otimes \tilde{\phi}^{i'}.
\ee 
To recast this observation into the  language  used for $\mf{h}(\N_L, \N_R, \mathds{D})$, we summarise here the corresponding division algebraic characterisation of the $(D, \N)$ super Yang-Mills multiplet $(A_{\mu}, \lambda^{a}, \phi^{i})$:
(i)   The gauge potential $A_{\mu}$ is a $\mathfrak{sa}(\N, \mathds{D})$ singlet valued in $\R$. 
(ii)  The $\N$ gaugini $\lambda^{a}$ transform in the defining representation  of $\mathfrak{sa}(\N, \mathds{D})$ and are valued in $\mathds{D}^\N$.
(iii) The  scalars $\phi^{i}$ span a subspace  $\mathds{D}_{*}[\mathcal{N}]\subseteq \mathds{D}[\N]\cong\mathds{D}^\mathcal{N}\otimes\mathds{D}^\mathcal{N}$ since they are quadratic in the supersymmetry charges valued in $\mathds{D}^\N$.  Some brief remarks on $\mathds{D}_*$ are in order. Recall,  the spectrum of massless states can be constructed using a supercharge  annihilation operator  carrying the minimal spinor representation $S$ of $\mf{so}(2)\oplus\mf{so}(D-4)\subset\mf{so}(D-2)$ and  the defining representation $\mathds{D}^\mathcal{N}$ of $\mf{sa}(\mathcal{N}, \mathds{D})$, with $\mf{so}(2)$ charge $(-1/2)$.  Under $\mf{so}(2)\oplus\mf{so}(D-4)\subset\mf{so}(D-2)$ the little group vector state splits as 
\be
\R^{D-2}=\R_{\1}\oplus\R_{\sst{(-1)}}\oplus\R^{D-4}_{\0}. 
\ee
Acting twice on the  vacuum, defined as the positive grade  $\mf{so}(D-4)$  singlet of the vector,  yields the scalar states and the $D-4$ vector states with zero $\mf{so}(2)$ charge living in,
\be
\left[\wedge^2(S\otimes\mathds{D}^\mathcal{N})\right]_\R=\left[\wedge^2(S)\otimes\text{Sym}^2(\mathds{D}^\mathcal{N})+\text{Sym}^2(S)\otimes\wedge^2(\mathds{D}^\mathcal{N})\right]_\R,
\ee
where $]_\R$ denotes projection with respect to the appropriate real structure. The $D-4$ vector states  are in the $\R^{D-4}_{\0}$ representation of $\mf{so}(2)\oplus\mf{so}(D-4)$ and are necessarily singlets under $\mf{sa}(\mathcal{N}, \mathds{D})$. Conversely, the remaining states correspond to scalars, which  are necessarily singlets under $\mf{so}(D-4)$. Denoting by
$\varphi_*$
the projector onto the orthogonal compliment of $\R^{D-4}\subset\left[\wedge^2(S\otimes\mathds{D}^\mathcal{N})\right]_\R$ we define,
\be
\mathds{D}_*[\mathcal{N}] = \varphi_*\left[\wedge^2(S\otimes\mathds{D}^\mathcal{N})\right]_\R.
\ee
We do not adopt a ``really real'' basis for the scalars, wishing to keep $\mathds{D}$ structure of the algebras manifest. Consider $D=7, \mathcal{N}=1$ as an example. The supercharge  annihilation operator transforms as the $\rep{(2,2)}$ of $\mf{so}(3)\oplus\mf{sa}(1, \Q)\cong \mf{sp}(1)\oplus\mf{sp}(1)$. From
$
\wedge^2(\rep{2,2})=\rep{(3,1)+(1,3)}
$
we obtain vector states in $\rep{(3,1)}\cong\R^3$ and scalar states in $\rep{(1,3)}\cong\text{Im}\Q\cong\mathds{D}_*[1]$.  A detailed  description of the   $\mathds{D}_{*}[\mathcal{N}]$ spaces is given in \autoref{scalartab} of  \autoref{Scalar terms in SYM}. Interestingly, the subspace $\mathds{D}_{*}[\mathcal{N}]$ (up to an algebra automorhpism) may be concisely characterised as   $\mathds{D}[\mathcal{N}]$ matrices satisfying a set of ``Clifford-like'' conditions:
\begin{subequations}
\begin{align}
\label{cond_conj_fin}
MN^{\sigma}+NM^{\sigma}&=(-1)^\mathcal{D}(m n^*+n m^*)\mathds{1},\\
\label{cond_dagger_fin}
MM^\dagger&=|m|^2\mathds{1},
\end{align}
\end{subequations}
with $m, n\in \mathds{D}$ and $\mathcal{D}=\dim \mathds{D}$ being the dimension of the spinor representation in $D-2$, as given in \autoref{tab:AD}. We have defined an involution  $\sigma:\mathds{D}\rightarrow \mathds{D}$, which if we regard $\R=\text{Span}_{\R}\{e_0\}$ and $\C=\text{Span}_{\R}\{e_0, e_1\}$ as subalgebras in $\Q=\text{Span}_{\R}\{e_0, e_1, e_2, e_3\}$ can be defined by conjugation with respect to $e_2$, 
\be
m^\sigma=e_2me_{2}^{*}
\ee
or equivalently 
\be
\label{define_stargen}
e_{a}^{\sigma}=(-1)^ae_a,\qquad a=0,\ldots , 3.
\ee
Note, the quaternions are isomorphic to the $D=2+0$ Clifford algebra, $\Q\cong\Cliff(\R^2)$, and $\sigma$ is the \emph{canonical automorphism} on $\Cliff(\R^2)$ \cite{Daboul:1999xv}.

Each component of $\mathfrak{p}$ decomposed with respect to $\mathfrak{int}(\N_L, D)\oplus\mathfrak{int}(\N_L, D)$  then has a direct $left\otimes right$ origin \eqref{eq:scalars} expressed in terms of the above representation spaces:

\begin{enumerate}
\item $A_{\mu}\otimes\tilde{A}_{\nu}$:  The scalars originating from $A_L\otimes A_R$ belong to $\R_L\otimes\R_R\cong\mathfrak{so}(1,1)$. In $D=4$, there is an extra  $\R_L\otimes\R_R$ term originating from the dualisation  $B_{\mu\nu}\rightarrow\phi$. This contributes to $\mathfrak{p}$ a term given by
\be
\R_L\otimes\R_R+i\delta_{D,4}\R_L\otimes\R_R.
\ee
\item $\lambda^{a}\otimes \tilde{\lambda}^{a'}$: The scalars originating from $\lambda_L\otimes\lambda_R$ contribute a term given by
\be
\mathds{D}^{\N_L}\otimes\mathds{D}^{\N_R}\cong\mathds{D}[\mathcal{N}_L, \mathcal{N}_R].\ee

\item $\phi^{i}\otimes \tilde{\phi}^{i'}$: The scalars  originating from $\phi_L\otimes\phi_R$ contribute a term given by
\be
 \mathds{D}_{*}[\mathcal{N}_L]\otimes\mathds{D}_{*}[\mathcal{N}_R].\ee
\end{enumerate}
Bringing  these elements together, we conclude that in total  $\mathfrak{g}$ as a vector space is given by:
 \be\label{eq:Galg2}
\mathfrak{g}(\mathcal{N}_L+\mathcal{N}_R,D)
=\mathfrak{h}(\mathcal{N}_L+\mathcal{N}_R,D)+\mathds{D}_{*}[\mathcal{N}_L]\otimes\mathds{D}_{*}[\mathcal{N}_R]+\mathds{D}[\mathcal{N}_L, \mathcal{N}_R]+\R_L\otimes\R_R+i\delta_{D,4}\R_L\otimes\R_R.
\ee 
In \autoref{eq:Gcomms} we present a set of commutators which define a Lie algebra structure on \eqref{eq:Galg2}, giving precisely the algebras of the generalised U-duality pyramid in \autoref{fig:Galldims}. To describe the complete set of commutators  we use the left/right form of $\mathfrak{h}\subset\mathfrak{g}$ given in \eqref{eq:Halg2},
 \be\label{eq:Galg3}
\mathfrak{h}(\mathcal{N}_L+\mathcal{N}_R,D)
=\Big[\underbrace{\mathfrak{sa}(\mathcal{N}_L, \mathds{D})\oplus\mathfrak{sa}(\mathcal{N}_R, \mathds{D})\oplus\delta_{D,4}\mathfrak{u}(1)}_{\bar{\mathfrak{h}}(\mathcal{N}_L+\mathcal{N}_R,D)}+\mathds{D}[\mathcal{N}_L, \mathcal{N}_R]_{c}\Big]\oplus \mathfrak{p}_L\oplus\mathfrak{p}_R
\ee 
where we have distinguished the compact $\mathds{D}[\mathcal{N}_L, \mathcal{N}_R]_{c}\subset \mathfrak{h}$ and the non-compact $\mathds{D}[\mathcal{N}_L, \mathcal{N}_R]_{nc}\subset \mathfrak{p}$. The non-trivial commutators amongst the compact generators have been given in \eqref{eq:racomms}.   
\newcolumntype{N}{>{$}l<{$}}
\newcolumntype{P}{>{$}r<{$}}
\begin{table}[h]
\center
\begin{tabular}{PNNMNNNNNNNNMMMMMMMMMMMMMMMMMMMMMM}
\bar{\mathfrak{h}}(\mathcal{N}_L+\mathcal{N}_R,D)&\times &\mathds{D}[\mathcal{N}_L, \mathcal{N}_R]_{nc} &\rightarrow& \mathds{D}[\mathcal{N}_L, \N_R]_{nc}\\[4pt]
X_L\oplus X_R&\otimes &P&\mapsto&X_LP-PX_R\\[14pt]

\bar{\mathfrak{h}}(\mathcal{N}_L+\mathcal{N}_R,D)&\times& \mathds{D}_{*}[\mathcal{N}_L]\otimes\mathds{D}_{*}[\mathcal{N}_R] &\rightarrow& \mathds{D}_{*}[\mathcal{N}_L]\otimes\mathds{D}_{*}[\mathcal{N}_R]\\[4pt]
X_L\oplus X_R&\otimes&m\otimes p&\mapsto&(X_Lm-mX_L^{\sigma})\otimes p+m\otimes(X_Rp-pX_R^{\sigma})\\[12pt]

\mathds{D}[\mathcal{N}_L, \mathcal{N}_R]_{c} &\times& \R+ i\delta_{D,4}\R&\rightarrow&  \mathds{D}[\N_L, \N_R]_{nc}\\[4pt]
M&\otimes&\alpha&\mapsto&\alpha M\\[14pt]

 \mathds{D}[\mathcal{N}_L, \mathcal{N}_R]_{c} &\times& \mathds{D}[\mathcal{N}_L, \mathcal{N}_R]_{nc} &\rightarrow& \mathds{D}_{*}[\mathcal{N}_L]\otimes\mathds{D}_{*}[\mathcal{N}_R]+ \R+ i\delta_{D,4}\R\\[4pt]
M&\otimes&P&\mapsto&\varphi_*(M,P)+ \tr(MP)\\[14pt]

 \mathds{D}[\mathcal{N}_L, \mathcal{N}_R]_{c} &\times& \mathds{D}_{*}[\mathcal{N}_L]\otimes\mathds{D}_{*}[\mathcal{N}_R] &\rightarrow& \mathds{D}[\N_L, \N_R]_{nc}\\[4pt]
M&\otimes&m \otimes p&\mapsto&\frac{4}{3}mMp^{\sigma} \\[14pt]

 \mathds{D}[\mathcal{N}_L, \mathcal{N}_R]_{nc} &\times& \R+ i\delta_{D,4}\R&\rightarrow& \mathds{D}[\N_L, \N_R]_{c}\\[4pt]
P&\otimes&\alpha&\mapsto&\alpha P  \\[14pt]

 \mathds{D}[\mathcal{N}_L, \mathcal{N}_R]_{nc} &\times&  \mathds{D}[\mathcal{N}_L, \mathcal{N}_R]_{nc}&\rightarrow& \bar{\mathfrak{h}}(\mathcal{N}_L+\mathcal{N}_R,D)\\[4pt]
P&\otimes&Q&\mapsto&(PQ^\dagger - QP^\dagger)\oplus (PQ^\dagger - QP^\dagger) \\[14pt]

 \mathds{D}[\mathcal{N}_L, \mathcal{N}_R]_{nc} &\times&  \mathds{D}_{*}[\mathcal{N}_L]\otimes\mathds{D}_{*}[\mathcal{N}_R] &\rightarrow&  \mathds{D}[\N_L, \N_R]_{c}\\[4pt]
P&\otimes&m \otimes p&\mapsto&\frac{4}{3}mPp^{\sigma}  \\[14pt]

\mathds{D}_{*}[\mathcal{N}_L]\otimes\mathds{D}_{*}[\mathcal{N}_R]  &\times&  \mathds{D}_{*}[\mathcal{N}_L]\otimes\mathds{D}_{*}[\mathcal{N}_R] &\rightarrow& \bar{\mathfrak{h}}(\mathcal{N}_L+\mathcal{N}_R,D)\\[4pt]
m \otimes p&\otimes&n \otimes q&\mapsto&(mn^\dagger-nm^\dagger )\tr(pq^\dagger)\oplus (pq^\dagger-qp^\dagger) \tr(mn^\dagger)\\[14pt]
\end{tabular}
\caption{Commutators of $\mathfrak{g}(\mathcal{N}_L+\mathcal{N}_R,D)$-algebra given in \eqref{eq:Galg2}.\label{eq:Gcomms}}
\end{table}

The remaining non-trivial commutators  are  given in \autoref{eq:Gcomms}. Through out matrix multiplication is used except for
\be
[\mathds{D}[\mathcal{N}_L, \mathcal{N}_R]_{c},  \mathds{D}[\mathcal{N}_L, \mathcal{N}_R]_{nc}]\subseteq \mathds{D}_{*}[\mathcal{N}_L]\otimes\mathds{D}_{*}[\mathcal{N}_R]+ \R+ i\delta_{D,4}\R,
\ee
where we have introduced the natural extension of the   projector $\varphi_*$,
\be
\hat{\varphi}_*: [\mathds{D}^m \otimes \mathds{D}^n]\otimes [\mathds{D}^m \otimes \mathds{D}^n]\rightarrow \mathds{D}_*[m]\otimes\mathds{D}_*[n].
\ee
Note, for the sake of brevity we have  reincorporated the $D=4, \mf{u}(1)$ factor back into $X_L$ and $X_R$, which therefore have equal and opposite traces. Moreover, leaving aside $D=3$ for the moment, the only non-vanishing $\alpha\oplus\beta\in \mf{p}_L\oplus\mf{p}_R$ occur in  $D=4$,  the $\mf{u}(1)$ factors of   $\N=2,1$. See \autoref{R-symmetries in all dimensions table}.   Simply regarding $X_{L/R}$ as tracefull generators belonging to $\mf{a}(\N_{L/R})$ automatically accounts for their action. 

In three dimensions the formula can be simplified by ``dualising'' the $A_L\otimes A_R$ contributions into $\phi_L\otimes \phi_R$ terms. We  no longer have the $\R_L\otimes\R_R$ term from tensoring the gauge fields, it is combined into a second $\R[\N_L, \N_R]_{nc}$ factor resulting in the simplified $D=3$ formula, 
\be
\mf{g}(\N_L+ \N_R) = \mf{h}(\N_L, \N_R)+2\R[\N_L, \N_R]_{nc},
\ee
together with a simplified set of commutation relations \cite{Anastasiou:2013hba}.

We conclude this discussion by relating this perspective back to the $\alg=\R, \C, \Q, \Oct$ framework developed in our previous work \cite{Borsten:2013bp, Anastasiou:2013cya, Anastasiou:2013hba}. 
Each super Yang-Mills theory comes from reducing the fundamental $D=3,4,6,10$, $\mathcal{N}=1$ multiplet, and hence can be thought of in terms of those theories. This naturally associates $\R, \C, \Q,\Oct$ with each super Yang-Mills theory, according as to whether it   came from the $D=3,4,6,10$ theory, respectively. Equivalently, if one does not want to talk in terms of dimensional reduction,  the associated division algebra is just $\alg_n$ with $n=Q/2$, where $Q$ is the number of real supercharge components. The fermions can then be arranged into a single $\alg_n$ element. For example, in $D=4, \N=4$ the four complex spinors become one octonion.  The vectors and scalars, on the other hand, inhabit two orthogonal subspaces of $\alg_n$: the vector subspace is of course isomorphic to $\R^{(D-2)}$ while  the scalars span the complementary subspace  $\R^{(n-(D-2))}$.

 Then $\Spin(D-2)\subset \text{Cliff}_0(D-2)\cong\text{Cliff}(D-3)$ acts on the $\alg_n$-valued fermions via left multiplication by the $(D-3)$-dimensional imaginary part of the $\R^{(D-2)}$ subspace of $\alg_n$: in $D=4$ it would be generated by left-multiplication by the single element $e_1$, in $D=5$ it would be $e_1, e_2$, and so on. This is used to generate the spacetime little group transformations, and defines an isomorphism between $\alg_n$ and ($S_{D-2})^\N$, where $S_{D-2}$ is the spinor representation of $\Spin(D-2)\subset\text{Cliff}_{0(D-2)}$  and  $\mathcal{N}$ is the number of spinor supercharges (or number of fermions).

Left-multiplication by the complementary scalar subspace $\R^{(n-(D-2))}$ of $\alg_n$ defines a set of linear maps on $(S_{D-2})^\N$ isomorphic to a subset of $\N\times \N$ $\mathds{D}$-valued matrices, i.e. $\mathds{D}_*[N]$. For example, in $D=4$ left-multiplication of an octonion by the six basis elements $e_2, e_3, e_4, e_5, e_6, e_7$ defines six linear maps isomorphic to a set of $4\times4$ complex matrices (where the complex structure is given by left-multiplication of $e_1$), which is precisely those of $\mathds{D}_*[\N]$. Hence the result:
\be
\mathds{D}_*[\N]= \{L_x\,|\,  x \in \R^{(n-(D-2))}\subset \alg_n\}
\ee
where $L_x: \alg_n\rightarrow\alg_n$ denotes left-multiplication  by $x$.

\section{Discussion}

We have shown that the U-duality algebras $\mf{g}$ for all supergravity multiplets  obtained by tensoring two super Yang-Mills multiplets in $D\geq 3$ can be written in a single formula with three arguments, $\mf{g}(\N_L+ \N_R, \mathds{D})$. The formula relies on the link between the three associative normed division algebras, $\R, \C, \Q$, and the representation theory of classical Lie algebras. The formula is symmetric under the interchange of $\N_L$ and $\N_R$ and provides another ``matrix model'', in the sense of Barton and Sudbery \cite{Barton:2003}, for the exceptional Lie algebras. In this language the compact subalgebra $\mf{h}(\N_L+ \N_R, \mathds{D})$ has a simple form which makes the $left\otimes right$ structure clear. The non-compact $\mf{p}=\mf{g}-\mf{h}$ generators are obtained directly by examining  the division algebraic representations carried by those left/right states that produce the scalar fields of the corresponding supergravity multiplets. 

Note, we are therefore implicitly assuming that the tensor product always gives supergravities with scalars parametrising a symmetric coset space. The only possible exception to this rule is given by $\N_{L}=\N_R=1$. When there is  a possible ambiguity in the coupling of the scalars it is resolved by the  structure of the left and right symmetry algebras. For example, in $D=4$ the $\N_{L}=\N_R=1$  scalar coset manifold,  
\be
\frac{\Un(1,2)}{\Un(1)\times\Un(2)},
\ee
is the unique possibility consistent with the left and right super Yang-Mills  data.

This procedure  gives all supergravity algebras with more than half-maximal supersymmetry. These cannot couple to matter, as reflected by the squaring procedure where only the fields of the supergravity multiplet are produced. However, for half-maximal and below, one can couple the theory to matter multiplets (vector or hyper). This does indeed happen when one squares; the fields obtained arrange themselves in the correct number of vector or hypermultiplets such that we fill up the entries of the generalised  pyramid.

Theories with more general matter content  do not naturally live in our  pyramid, mainly because they lack an obvious division algebraic description. For example, the $STU$ model \cite{Duff:1995sm} is given by $\N=2$ supergravity in four dimensions coupled to three vector multiplets, while the entry for $\N=2$ in our pyramid  necessarily comes coupled to a single hypermultiplet. Can squaring accommodate more general matter couplings? All   factorized orbifold projections (as defined in \cite{Chiodaroli:2014xia}) of $\N = 8$ supergravity can be obtained from the tensor product of the corresponding left and right orbifold projections of $\N=4$ super Yang-Mills multiplets \cite{Chiodaroli:2014xia}. This includes a large, but still restricted, class of matter coupled supergravities with specific U-dualities.

Theories coupled to an arbitrary number of vector multiplets  can be obtained by tensoring   a supersymmetric multiplet with  a conveniently chosen collection of bosonic fields. In particular, here we consider an $\N_R=0$ multiplet with a single gauge potential and $n_V$ scalar fields. The symmetries of the resulting supergravity multiplet are determined by the global symmetries postulated for the  $\N_R=0$ multiplet. We consider  the  simplest case where the $n_V$ scalar fields transform in the vector representation  of a global $\SO(n_V)$. Following the procedure used to construct the generalised pyramid this uniquely fixes the global symmetries of the resulting  supergravity multiplet and therefore, implicitly, the structure of the matter couplings. This idea is developed in the following section. We summarise the results\footnote{Note that we have excluded  $\mathcal{N}=1$ theories in four dimensions. It is not possible to obtain $\mathcal{N}=1$ supergravity coupled to only vector multiplets by squaring since one always obtains at least one chiral multiplet when tensoring $\mathcal{N}=1$ SYM with a non-supersymmetric multiplet. The same applies to $\mathcal{N}=(1,0)$ supergravity in 6 dimensions. These theories are interesting in their own right and will be analysed in forthcoming  work \cite{Anastasiou:2015aa}.} in \autoref{General theories coupled to matter in D=3,4,5,6}.

 Note, more generally these examples of factorisable $\N\leq 4$ matter coupled supergravities are also physically interesting. In particular, they can be used to  shed light on the  UV divergences appearing in $\N\leq 4$ supergravity theories. Indeed, this double-copy construction of additional $\N=4$ vector multiplets was used in \cite{Carrasco:2013ypa} to isolate the effect of the Marcus anomaly  on $\N=4$ supergravity scattering amplitudes. Moreover, in \cite{Bern:2013uka} the dependence on $n_V$ was used to relate this anomaly to the 4-loop divergences appearing in these theories\footnote{We thank one of our referees for bringing these developments to our attention.}.

\subsection{$[\N_L]_V \times[\N_R=0]$ tensor products}\label{sec:GeneralisedTensorProducts}

\newcolumntype{M}{>{$}c<{$}}
\newcolumntype{N}{>{$}l<{$}}
\begin{table}[h]
\center
\scriptsize
\begin{tabular}{M|M|M|M|M}
\hline
\hline
&&&&\\
theory& squaring\ formula &R_L&R_R& \frac{G}{H}  \\[8pt]
\hline
&&&&\\
D=3     &&&&  \\[8pt]
\hline
&&&&\\
(\mathcal{N}=8)_{SuGra}+n_v(\mathcal{N}=8)_{vector} &(\mathcal{N}=8)_{V}\times[n_V\phi]  &\Spin(8)&\SO(n_V)& \frac{\SO(8,n_V)}{\SO(8)\times \SO(n_V)}   \\[8pt]
(\mathcal{N}=4)_{SuGra}+n_v(\mathcal{N}=4)_{vector} &(\mathcal{N}=4)_{V}\times[n_V\phi]   &\Spin(4)&\SO(n_V)& \frac{\SO(4,n_V)}{\SO(4)\times \SO(n_V)}   \\[8pt]
(\mathcal{N}=2)_{SuGra}+n_v(\mathcal{N}=2)_{vector} &(\mathcal{N}=2)_{V}\times[n_V\phi]  &\Spin(2)&\SO(n_V)&\frac{\SO(2,n_V)}{\SO(2)\times \SO(n_V)}  \\[8pt]
(\mathcal{N}=1)_{SuGra}+n_v(\mathcal{N}=1)_{vector} &(\mathcal{N}=1)_{V}\times[n_V\phi]  &\varnothing&\SO(n_V)&\frac{\SO(1,n_V)}{ \SO(n_V)}  \\[8pt]
\hline
&&&&\\
D=4  &&&& \\[8pt]
\hline
&&&&\\
(\mathcal{N}=4)_{SuGra}+n_v(\mathcal{N}=4)_{vector} &(\mathcal{N}=4)_{V}\times[A_\mu+n_V\phi]   &\SU(4)&\SO(n_V)& \frac{\SO(6,n_V)}{\SO(6)\times \SO(n_V)}\times \frac{\SL(2)}{\SO(2)}  \\[8pt]
(\mathcal{N}=2)_{SuGra}+n_v(\mathcal{N}=2)_{vector} &(\mathcal{N}=2)_{V}\times[A_\mu+(n_V-1)\phi]   &\Un(2)&\SO(n_V-1)& \frac{\SU(2)\times \SO(2,n_V-1)}{U(2)\times \SO(n_V-1)}\times \frac{\SL(2)}{\SO(2)}   \\[8pt]
\hline
&&&&\\

D=5  &&&& \\[8pt]
\hline
&&&&\\

(\mathcal{N}=2)_{SuGra}+n_v(\mathcal{N}=2)_{vector} &(\mathcal{N}=2)_{V}\times[A_\mu+n_V\phi]   &\Sp(2)&\SO(n_V)& \frac{\SO(5,n_V)}{\SO(5)\times \SO(n_V)}\times \Orth(1,1)   \\[8pt]
(\mathcal{N}=1)_{SuGra}+n_v(\mathcal{N}=1)_{vector} &(\mathcal{N}=1)_{V}\times[A_\mu+(n_V-1)\phi]   &\Sp(1)&\SO(n_V-1)& \frac{\Sp(1)\times \SO(1,n_V-1)}{\Sp(1)\times \SO(n_V-1)}\times \Orth(1,1)    \\[8pt]
\hline
&&&&\\
D=6  &&&& \\[8pt]
\hline
&&&&\\

\begin{array}{c}(\mathcal{N}=(1,1))_{SuGra}\\ +n_v(\mathcal{N}=(1,1))_{vector}\end{array} &(\mathcal{N}=(1,1))_{V}\times[A_\mu+n_V\phi]  &\Sp(1)\times \Sp(1)&\SO(n_V)& \frac{\Orth(4,n_V)}{\SO(4)\times \Orth(n_V)}\times \Orth(1,1)    \\[8pt]
\begin{array}{c}(\mathcal{N}=(2,0))_{SuGra}\\ +n_T(\mathcal{N}=(2,0))_{tensor}\end{array} &(\mathcal{N}=(2,0))_{tensor}\times[B_{\mu\nu}^-+n_T\phi]  & \Sp(2)&\SO(n_T)& \frac{\Orth(5,n_T)}{\SO(5)\times \Orth(n_T)}\\[8pt] 
\hline
\hline
\end{tabular}
\caption{Matter  coupling  in $D=3,4,5,6$}\label{General theories coupled to matter in D=3,4,5,6}
\end{table}
Note that the general form for the maximally compact subgroups in the cosets given in \autoref{General theories coupled to matter in D=3,4,5,6} is 
\be
H=R_L\otimes R_R\otimes {\delta_{D, 4}}\SO(2).
\ee
This is just the form,
\be
\mathfrak{h}(\mathcal{N}_L+\mathcal{N}_R,D)=\mathfrak{int}(\mathcal{N}_L,D)\oplus\mathfrak{int}(\mathcal{N}_R,D)\oplus\delta_{D,4}\mathfrak{u}(1)+\mathds{D}[\mathcal{N}_L, \mathcal{N}_R],
\ee
 appearing in the generalised  pyramid formula \eqref{eq:Halg2} with $\mathds{D}[\mathcal{N}_L, \mathcal{N}_R=0]=\varnothing$.   This is entirely consistent with the logic of the construction; we previously  identified $\mathds{D}[\mathcal{N}_L, \mathcal{N}_R]$ with the generators $Q_L\otimes Q_R$, which are clearly absent when $\N_R=0$.

The non-compact generators are also determined  following the logic of the generalised pyramid presented in  \autoref{sec:Gsym}, but now with only two   scalar terms: $A_{\mu}\otimes\tilde{A}_{\nu}$ and $\phi^{i}\otimes \tilde{\phi}^{i'}$, where $\tilde{\phi}^{i'}$ are the $n_V$ scalars   transforming as a vector of $\SO(n_V)$. 

As an example, take half-maximal supergravity in five dimensions coupled to $n_V$ vector multiplets. We obtain the field content by tensoring the maximal $\mathcal{N}=2$ super Yang-Mills  multiplet (with R-symmetry  $\Sp(2)$) and a non-supersymmetric multiplet consisting of a gauge field and $n_V$ scalars transforming in the vector representations  of $\SO(n_V)$, denoted $\rep{n}_V$:
\be\label{eq:D5nvectors_ex}
\begin{array}{c|ccccccc}
\otimes
&
 \begin{array}{c} \tilde{A}_\mu\\(\rep{3}; \rep{1})\end{array}&  \begin{array}{c} \tilde{\phi} \\(\rep{1}; \rep{n_V})\end{array}\\
\hline
A_\mu\;\;(\rep{3}; \rep{1}) &(\rep{5}; \rep{1,1})+ (\rep{3}; \rep{1,1})+(\rep{1}; \rep{1,1})&(\rep{3}; \rep{1,n_V})\\
\lambda\;\;(\rep{2}; \rep{4})&(\rep{4}; \rep{4,1})+(\rep{2}; \rep{4,1})& (\rep{2}; \rep{4,n_V})\\
 \phi \;\; (\rep{1}; \rep{5})&(\rep{3}; \rep{5,1})&(\rep{1}; \rep{5,n_V})
 \end{array}
\ee
We therefore find,
\be
\mf{h}=\mf{sp}(2)\oplus \mf{so}(n_V), 
\ee
and, from \eqref{eq:D5nvectors_ex}, 
\be
\mf{g}\ominus\mf{h}=(\rep{5},\rep{n_V})\oplus(\rep{1},\rep{1}).
\ee
Using the commutators which follow uniquely from the transformation properties of left and right states we have
\be
\mf{g}=[\mf{sp}(2)\oplus \mf{so}(n_V) +(\rep{5},\rep{n_V})]\oplus (\rep{1},\rep{1})\cong \mf{so}(5, n_V)\oplus\mf{so}(1,1).
\ee
This procedure applied in $D=3,4,5,6$ yields \autoref{General theories coupled to matter in D=3,4,5,6}. Note, for   $D=4$, $\mathcal{N}=2$ and   $D=5$, $\mathcal{N}=1$ the $\SU(2)$ and $\Sp(1)$ factors, respectively, drop out of the $G/H$ coset. We see that the cosets admit a concise alternative description:
\be
\frac{G}{H}\cong\frac{\SO(\#_{\phi_L},\#_{\phi_R})}{\SO(\#_{\phi_L})\times \SO(\#_{\phi_L})}\times \mathcal{M}_{A_L\times A_R}.
\ee
where $\#_{\phi_{L/R}}$ is the number of scalars in the left and right multiplets we are tensoring and $\mathcal{M}_{A_L\times A_R}$ is the coset parametrised by the scalars obtained from tensoring the gauge fields. It is given by $\varnothing$ in $D=3$, since the gauge fields (in the free theory) have been dualised to scalars, $\SL(2)/\SO(2)$ in $D=4$ where we have two such scalars, and $\Orth(1,1)$ in $D=5,6$, where we have one.  

In some cases we  reproduce cosets appearing in the generalised pyramid. For example, in $D=4$ $[\N=2]_V\times[\N=2]_V$ and  $[\N=4]_V\times [\N=0, n_V=2]$ both yield $\N=4$ supergravity coupled to two vector multiplets with coset $[\SL(2)\times \SO(6,2)]/[\SO(2)^2\times \SO(6)]$.  However, despite their common coset the two resulting theories have important  structural differences when interpreted as truncations of $D=4, \N=8$ supergravity. In particular, the $\SL(2)$ S-duality subgroup in $E_{7(7)}$ can be directly identified with  the $\SL(2)$ factor in $\SL(2)\times \SO(6,2)\subset E_{7(7)}$ for $[\N=4]_V\times [\N=0, n_V=2]$ whereas for 
$[\N=2]_V\times[\N=2]_V$ it must be identified with an $\SL(2)$ subgroup of the $\SO(6,2)$ factor, as explained in \cite{Anastasiou:2013hba}. In both cases    the $8+8$ gauge potentials and their duals transform as the $(\rep{2, 8})$ of $\SL(2)\times \SO(6,2)$. Embedding the  $[\N=2]_V\times[\N=2]_V$ theory in $\N=8$ supergravity these $8+8$ potentials and dual potentials are evenly split between the  NS-NS and RR sectors, implying that the $\SL(2)$ factor mixes NS-NS and RR potentials and therefore cannot be identified with S-duality. Instead, the S-duality $\SL(2)_S$ is contained in the $\SO(6, 2)$ component:
\be
\begin{split}
\SL(2)\times\SO(6,2)&\supset \SL(2)\times\SL(2)_S\times\SL(2)\times\SU(2),\\
(\rep{2,8})&\rightarrow \underbrace{(\rep{2}, \rep{2}_S,\rep{2,1,1})}_{\text{NS-NS}}+\underbrace{(\rep{2}, \rep{1}_S,\rep{1,2,2})}_{\text{RR}}.
\end{split}
\ee

On the other hand, the $\N=4$ supergravity coupled to two vector multiplets obtained from $[\N=4]_V\times [\N=0, n_V=2]$ can be consistently embedded in the NS-NS sector of $\N=8$ supergravity alone: all eight gauge potentials correspond to NS-NS states. In this scenario, the $\SL(2)$ factor in the U-duality group can be identified as  the S-duality $\SL(2)\in E_{7(7)}$:
\be
\begin{aligned}
\SL(2)\times \SO(6,2)&\cong \SL(2)_S\times \SO(6,2), \\
(\rep{2},\rep{8})&\equiv (\rep{2}_S,\rep{8}).
\end{aligned} 
\ee

\section*{Acknowledgments}

We would like to thank  MJ Duff and A Marrani for very many useful and instructive conversations on the symmetries of supergravity. The work of LB was supported by a Sch\"{o}dinger Fellowship and an Imperial College Junior Research Fellowship. AA, MJH, and SN received support from STFC and EPSRC.

\appendix

\section{Scalars in $\mathds{D}[\mathcal{N}]$}\label{Scalar terms in SYM}
In \autoref{scalartab} we have  listed, for each dimension $D$ and number of supersymmetry charges $Q$, the R-symmetry algebra $\mf{a}(\N, \mathds{D})$, the $\mf{a}(\N, \mathds{D})$-representation $\rep{n}$ carried by the scalar fields and the corresponding representation space $\mathds{D}_*[\N]\subseteq\mathds{D}[\N]$.

\newcolumntype{M}{>{$}c<{$}}
\begin{table}[h]
\center
\begin{tabular}{M|M|M|MMMMMM}
\hline
\hline
D/Q &16&8&4\\
\hline
10&\begin{array}{c} \mf{a}((1,0), \R^+\oplus\R^-)\cong\varnothing\\ \varnothing\subseteq \R^+[1]\end{array} &-&-\\[12pt]

9&\begin{array}{c} \mf{a}(1, \R)\cong \varnothing\\ \rep{1}\cong\R\subseteq \R[1]\end{array} &-&-\\[12pt]

8&\begin{array}{c} \mf{a}(1, \C)\cong\mf{u}(1)\\ (+2)+(-2)\cong\C\subseteq \C[1]\end{array} &-&-\\[12pt]

7& \begin{array}{c} \mf{a}(1, \Q)\cong\mf{sp}(1)\\ \rep{3}\cong\text{Im}\Q\subseteq \Q[1]\end{array} &-&-\\[12pt]

6& \begin{array}{c} \mf{a}((1,1), \Q^{+}\oplus\Q^{-})\cong\mf{sp}(1)\oplus\mf{sp}(1)\\ \rep{(2,2)}\cong\Q\subseteq \Q^+[1]\oplus\Q^-[1]\end{array} &\begin{array}{c} \mf{a}((1,0), \Q^{+}\oplus\Q^{-})\cong\mf{sp}(1)\\ \varnothing\subseteq\Q^+[1]\end{array}&-\\[12pt]

5& \begin{array}{c} \mf{a}(2, \Q)\cong\mf{sp}(2)\\ \rep{5}\cong\J_{2}^{0}(\Q)\subseteq \Q[2]\end{array} &\begin{array}{c} \mf{a}(1, \Q)\cong\mf{sp}(1)\\ \rep{1}\cong\text{Re}\Q\subseteq \Q[1]\end{array}&-\\[12pt]

4& \begin{array}{c} \mf{a}(4, \C)\cong\mf{u}(4)\\ \rep{6}_{0}\cong\wedge^{2}_{*}\C^4\subseteq \C[4]\end{array} &\begin{array}{c} \mf{a}(2, \C)\cong\mf{u}(2)\\ \rep{1}_{2}+\rep{1}_{-2}\cong\C\cong\wedge^{2}\C^2\subseteq \C[2]\end{array}&\begin{array}{c} \mf{a}(1, \C)\cong\mf{u}(1)\\ \varnothing\cong\wedge^{2}\C\subseteq \C[1]\end{array} \\[12pt]
\hline
\hline
\end{tabular}
\caption{R-symmetry algebra $\mf{a}(\N, \mathds{D})$, the $\mf{a}(\N, \mathds{D})$-representation $\rep{n}$ carried by the scalar fields and the corresponding representation space $\mathds{D}_*[\N]\subseteq\mathds{D}[\N]$. Note, $\J_{2}^{0}(\Q)$ is the space of traceless $2\times 2$ Hermitian matrices over $\Q$.\label{scalartab}}
\end{table}
 The perhaps less familiar cases over $\Q$ are summarised  here.
An element  
$
X\in\mf{u}(2n)\cong\mf{a}(2n, \C)
$
can be written
\be
X= \left(\begin{array}{cccc}
 a   &  b   \\
 -b^\dagger& c
\end{array}
\right),\quad\text{where}\quad a, c \in \mf{a}(n, \C)\quad b\in \C[n],
\ee
where the commutators are the usual matrix commutators.
For $X$ in the subalgebra  \be\mf{usp}(2n):=\{X\in\mf{u}(2n)| X^T\Omega+\Omega X=0\}\ee
we have
\be
X^T\Omega+\Omega X=0\quad\Rightarrow\quad
X= \left(\begin{array}{cccc}
 a   &  b   \\
 -b^\dagger& a^*
\end{array}
\right),\quad\text{where}\quad  b \in \text{Sym}^2(\C^n).
\ee
The well-known Lie algebra isomorphism $\mf{usp}(2n)\cong\mf{sa}(n, \Q)$ then follows from the  standard algebra bijection  
\be\label{tau}
\tau : \mf{sa}_u(2, \C) \rightarrow \Q\quad \text{where}\quad \mf{sa}_u(2, \C):=\text{Span}_{\R}\{\mathds{1}, \mf{sa}(2, \C)\}\ee
given by
\be
 \left(
\begin{array}{cccc}
 a_0 +i a_1  & a_2  +  i a_3  \\
 -a_2  + i a_3 & a_0 - i a_1
\end{array}
\right)\mapsto a_0 e_0 +a_1 e_1 + a_2 e_2 + a_3 e_3\in \Q,
\ee
where,
\be
\tau(X^\dagger)=\tau(X)^*,\quad\tau(X^*)=e_{2}\tau(X)e_{2}^{*},\quad\tau(X^T)=e_{2}\tau(X)^*e_{2}^{*}.
\ee
We can rewrite this using the notation introduced in \eqref{define_stargen} as:
\be
\label{useful_relations_tau}
\tau(X^\dagger)=\tau(X)^*,\quad\tau(X^*)=\tau(X)^{\sigma},\quad\tau(X^T)=(\tau(X)^*)^{\sigma}.
\ee

To set-up the  family of isomorphisms $\theta_{n}:\mf{usp}(2n)\rightarrow \mf{sa}(n, \Q)$ we introduce two linear maps, $\theta_{n}:=\tau_n\circ S_n$. First,  $S_n$ is a special orthogonal similarity transformation,
\be 
\label{rotate_X_appendix}
S_n:\C[2n]\rightarrow \C[2n];\quad X\mapsto \mathbf{S}_n{X}\mathbf{S}_{n}^{T},\qquad
\mathbf{S}_n\in \SO(2n),
\ee
organising $X$ into $2\times 2$ blocks $A_{ij}, i,j=1,2,\ldots,n$ such that $A_{ii}\in\mf{sa}(2, \C)$  and $A_{ji}=-A_{ij}^{\dagger}\in\mf{sa}_u(2, \C)$. Second,   $\tau_{n}: \R[n] \otimes_\R \mf{sa}_{u}(2, \C) \rightarrow \Q[n]$ is the natural extension of \eqref{tau} acting on $2\times 2$ blocks,
\be
\tau_{n}:  \left(
\begin{array}{cccc}
 A_{11}  &  A_{12}&\cdots &A_{1n}\\
 A_{21} & A_{22}&\cdots&A_{2n}\\
 \vdots & \vdots &\ddots&\vdots\\
  A_{n1}& A_{n2}&\cdots&A_{nn}\\
\end{array}
\right)\mapsto \left(
\begin{array}{cccc}
 \tau(A_{11})  &   \tau(A_{12})&\cdots & \tau(A_{1n})\\
  \tau(A_{21}) &  \tau(A_{22})&\cdots& \tau(A_{2n})\\
 \vdots & \vdots &\ddots&\vdots\\
   \tau(A_{n1})&  \tau(A_{n2})&\cdots& \tau(A_{nn})\\
\end{array}
\right).
\ee
Since   $A_{ii}\in\mf{sa}(2, \C)$  and $A_{ji}=-A_{ij}^{\dagger}\in\mf{sa}_u(2, \C)$ for  $S_n(X), X\in\mf{usp}(2n)$, 
$\tau_n\circ S_n (X)\in \mf{sa}(n, \Q)$. 

The similarity transformation $S_{n}$ is trivially a bijective  matrix algebra homomorphism and therefore also a Lie algebra isomorphism. Similarly, $\tau_n$ is trivially an algebra isomorphism since $\tau$ itself is an algebra isomorphism. Therefore its restriction to $\mf{ups}(2n)$ is  a Lie algebra isomorphism as the commutators are given by matrix commutators. Hence,
\be
\theta_{n}:\mf{usp}(2n)\rightarrow \mf{sa}(n, \Q)
\ee
is a Lie algebra isomorphism, such that
\be
\label{useful_relations_to_H}
\theta_n(X^\dagger)=\theta_n(X)^\dagger,\quad \theta_n(X^*)=\theta_n(X)^{\sigma},\quad \theta_n(X^T)=(\theta_n(X)^{\dagger})^{\sigma}.
\ee
As an example, consider the most relevant case of 
$
n=2
$:
\be
X=\left(
\begin{array}{cccc}
 ia & b& \alpha & \beta \\
 -b^* & ic & \beta & \delta \\
 -\alpha^* & -\beta^* & -ia & b^* \\
 -\beta^* & -\delta^* & -b & -ic \\
\end{array}
\right) \underset{S_2}{\longrightarrow}\left(
\begin{array}{cccc}
 ia & \alpha & b & \beta \\
   -\alpha^*& -ia & -\beta^* &  b^* \\
-b^* & \beta  &  ic &\delta \\
 -\beta^* & -b  &-\delta^* & -ic \\
\end{array}
\right) \underset{\tau_{2}}{\longrightarrow}\left(
\begin{array}{cccc}
 x &  z \\
  -z^*& y
\end{array}
\right) \in \mf{sa}(2, \Q).
\ee

In addition to the Lie algebra we need, in particular, the $\rep{5}$ of $\usp(4)\cong\mf{sa}(2, \Q)$.
The $\rep{6}$ of $\mathfrak{su}(4)$ is given by the space of  complex-self-dual 2-form  $m_{ab}\in\wedge^{2}\C^4, (m_{ab})^*=(\star m)^{ab}$. It can be written as a $4\times 4$ matrix
\be
m=
\begin{pmatrix}
0&\alpha&a&\beta\\
-\alpha&0&\beta^*&-a^*\\
-a&-\beta^*&0&\alpha^*\\
-\beta&a^*&-\alpha^*&0\\
\end{pmatrix}.
\ee
The action of $X_{a}{}^{b}\in \mf{su}(4)\cong\mf{sa}(4, \C)$ on $m_{ab}$ is given by
\be\label{eq:MX}
\begin{split}
 [X, m]_{ab}	&=X_{[a|}{}^{a'}m_{a'|b]}+X_{[b}{}^{b'}m_{a]b'}\\
			&=Xm-mX^*=Xm-mX^\sigma.
\end{split}
\ee
The Jacobi identity, 
\be
[X_1, [X_2, m]]+[X_2, [m, X_1]]+[m, [X_1, X_2]]=0,
\ee
holds for $[X_1, X_2]$ the usual matrix commutator, since $(X_1X_2)^*={X_1}^*{X_2}^*$ for $X_1, X_2\in \mf{sa}(n, \alg)$ (this is of course trivially true for $\alg=\R, \C$).

	Under $\su(4)\supset\usp(4)\cong\spa(2)$, 
\be
\rep{6}\rightarrow \rep{5}+\rep{1},
\ee
 where the $\rep{5}$ is a symplectic traceless complex-self-dual 2-form  $m_{ab}\in\wedge_{0}^{2}\C^4$,
\be
\Omega^{ab}m_{ab}=0, \qquad \Omega=\begin{pmatrix}
0&\mathds{1}\\
-\mathds{1}&0\\
\end{pmatrix}.
\ee
In terms of $m$ symplectic tracelessness implies $a\in\R$.
Applying  $S_2$ we obtain
\be\tilde{m}=\mathbf{S}m\mathbf{S}^T=
\left(
\begin{array}{cccc}
 0 & a & \alpha  & \beta  \\
 -a & 0 & -\beta ^* & \alpha ^* \\
 -\alpha  & \beta ^* & 0 & -a \\
 -\beta  & -\alpha ^* & a & 0 \\
\end{array}
\right)
\qquad
\text{where}
\qquad
\mathbf{S}=
\left(
\begin{array}{cccc}
 1 & 0 & 0 & 0 \\
 0 & 0 & 1 & 0 \\
 0 & 1 & 0 & 0 \\
 0 & 0 & 0 & 1 \\
\end{array}
\right)
\ee
so that
\be
\begin{split}
\theta_2(m)
&=\left(
\begin{array}{cccc}
 a e_2  & \alpha_0 + \alpha_1 e_1 + \beta_0 e_2 + \beta_1 e_3  \\
 -\alpha_0 - \alpha_1 e_1 + \beta_0 e_2 - \beta_1 e_3 & - a e_2
\end{array}
\right)\\
& =e_2  \left(
 \begin{array}{cccc}
 a   &b  \\
b^{*} & - a
\end{array}
\right), \qquad a\in \R, b\in \Q\\
& = e_2 A
\end{split}
\ee
where ${}^*$ now denotes the quaternionic conjugate and $A\in \J_{2}^{0}(\Q)$.  Applying $\theta_2$ to the commutator \eqref{eq:MX} we find
\be
\begin{split}
\theta_2([X, m]) &= \theta_2(X)\theta_2(m)-\theta_2(m)\theta_2(X)^\sigma\\
&=e_2( \theta_2(X)^\sigma A-A  \theta_2(X)^\sigma)\\
&= e_2 A'
\end{split}
\ee
where $A'\in \J_{2}^{0}(\Q)$. The  Jacobi identity follows trivially since $\theta_n$ is a matrix algebra homomorphism. 

For $D=6$ we can decompose with respect to the subalgebra $\mf{sa}(1, \Q)\oplus\mf{sa}(1, \Q)\subset\mf{sa}(2, \Q)$, 
\be
\begin{array}{cccccccccc}
\rep{10}&\rightarrow& \rep{(3,1)}+\rep{(1,3)}&+&\rep{(2,2)},\\
\left(
\begin{array}{cccc}
 x &  z \\
  -z^*& y
\end{array}
\right)
&\rightarrow&
\left(
\begin{array}{cccc}
 x &  0 \\
  0& y
\end{array}
\right)
&
+
&
\left(
\begin{array}{cccc}
 0 &  z \\
  -z^*& 0
\end{array}
\right).
\end{array}
\ee
Hence the $\rep{(2,2)}$ of $\mf{usp}(2)\oplus\mf{usp}(2)$ can be identified with $\Q$, where action of $x\oplus y\in \mf{sa}(1, \Q)\oplus\mf{sa}(1, \Q)$ on $z\in \Q$ is given by,
\be
[(x,y), z]=xz-zy.
\ee 
The $D=7$ subalgebra $\mf{sa}(1, \Q)$ is obtained by  identifying $x=y$. The $\rep{3}$ of $\mf{usp}(2)$ is then given by restricting $z$ to $\text{Im}\Q$,
\be
[(x,x), z]=xz-zx=xz-(xz)^*\in\text{Im}\Q.
\ee

%\bibliographystyle{utphys}
%\bibliography{Ref_Library}

\providecommand{\href}[2]{#2}\begingroup\raggedright\endgroup

\end{document}